\documentclass{pasj01}
\usepackage{url}
\usepackage[switch,mathlines]{lineno}
\Received{$\langle$reception date$\rangle$}
\Accepted{$\langle$acception date$\rangle$}
\Published{$\langle$publication date$\rangle$}

\RequirePackage{color}

\begin{document}

\title{Final results for mapping the Milky Way's stellar halo with blue
horizontal-branch stars selected from the Subaru Hyper Supreme-Cam Survey}

\author{Tetsuya~Fukushima\altaffilmark{1,2},
Masashi~Chiba\altaffilmark{2}, 
Mikito~Tanaka\altaffilmark{3}, 
Kohei~Hayashi\altaffilmark{4,2,5}, 
Daisuke~Homma\altaffilmark{1,2}, 
Sakurako~Okamoto\altaffilmark{1,6,7}, 
Yutaka~Komiyama\altaffilmark{3}, 
Masayuki~Tanaka\altaffilmark{1,6}, 
Nobuo~Arimoto\altaffilmark{7},
and Tadafumi~Matsuno\altaffilmark{8}
}

\altaffiltext{1}{National Astronomical Observatory of Japan, 2-21-1 Osawa, Mitaka, Tokyo 181-8588, Japan \\E-mail: {\it t.fukushima@astr.tohoku.ac.jp}}
\altaffiltext{2}{Astronomical Institute, Tohoku University, Aoba-ku, Sendai 980-8578, Japan}
\altaffiltext{3}{Department of Advanced Sciences, Faculty of Science and Engineering,
Hosei University, 184-8584 Tokyo, Japan}
\altaffiltext{4}{National Institute of Technology, Sendai College, Natori, Miyagi 981-1239, Japan}
\altaffiltext{5}{ICRR, The University of Tokyo, 5-1-5 Kashiwanoha, Kashiwa, Japan}
\altaffiltext{6}{The Graduate University for Advanced Studies, Osawa 2-21-1, Mitaka,
Tokyo 181-8588, Japan}
\altaffiltext{7}{Subaru Telescope, National Astronomical Observatory of Japan, 650 North A'ohoku Place,
Hilo, HI 96720, USA}
\altaffiltext{8}{Astronomisches Rechen-Institut, Zentrum f\"{u}r Astronomie der Universit\"{a}t Heidelberg, M\"{o}nchhofstr. 12-14, 69120 Heidelberg, Germany}


\maketitle

\begin{abstract}
We select blue-horizontal branch stars (BHBs) from the internal data release of the 
Hyper Suprime-Cam Subaru Strategic Program to reveal the global structure
of the Milky Way (MW) stellar halo. The data are distributed over $\sim 1,100$~deg$^2$ area in the range of
$18.5<g<24.5$~mag, so that candidate BHBs are detectable over a Galactocentric radius
of $r \simeq 36-575$~kpc. In order to select most likely BHBs by removing blue straggler stars and
other contaminants in a statistically significant manner, we develop and apply
an extensive Bayesian method, as described in \citet{Fukushima2019}. 
Our sample can be fitted to either a single power-law profile with an index of
$\alpha=4.11^{+0.18}_{-0.18}$ or a broken power-law profile with an index of
$\alpha_{\rm in}=3.90^{+0.24}_{-0.30}$ at $r$ below a broken radius of $r_{\rm b}=184^{+118}_{-66}$ kpc
and a very steep slope of $\alpha_{\rm out}=9.1^{+6.8}_{-3.6}$ at $r>r_{\rm b}$;
the statistical difference between these fitting profiles is small.
Both profiles are found to show prolate shapes having axial ratios of $q=1.47^{+0.30}_{-0.33}$ and 
$1.56^{+0.34}_{-0.23}$, respectively.
We also find a signature of the so-called "splashback radius" for the candidate BHBs, which can reach as large as $r \sim 575$~kpc,
although it is still inconclusive owing to rather large distance errors in this faintest end of the sample.
Our results suggest that the MW stellar halo consists of the two overlapping components:
the {\it in situ} inner halo showing a relatively steep radial density profile and
the {\it ex situ} outer halo with a shallower profile, being characteristic of a component
formed from accretion of small stellar systems.
\end{abstract}




\section{Introduction}

The current standard theory of structure formation based on $\Lambda$CDM models suggests that galaxies like the Milky Way (MW)
have been formed through hierarchical merging/accretion process from small to large scales \citep{White1978}.
While the memory of this past assembly history is weak or disappears in a bulge and disk, it is imprinted in a faint,
spatially extended stellar halo surrounding these brightest stellar components (e.g., see reviews,
\cite{Helmi2008,Ivezic2012,Feltzing2013,Bland-Hawthorn2014}). Individual merging debris of
small stellar systems like globular clusters and dwarf galaxies are present in the form of substructures in a stellar halo,
including stellar streams in the spatial distribution of stars and separate subcomponents in their velocity and/or phase-space
distribution (e.g., \cite{Helmi1999,Bullock2005,Cooper2010,Malhan2021,Suzuki2024}). The most dominant substructure in the MW stellar halo is made up
from the so-called Gaia-Sausage-Enceladus (GSE), which may have been originated from the merging event some 10~Gyr ago
\citep{Helmi2018,Belokurov2018}. Several other substructures have been discovered thanks to precise astrometric data by
{\it Gaia} satellite, and the cross-matching with photometry, spectroscopy and seismology data allows us to identify
the stellar population properties of these ancient progenitors (e.g., \cite{Helmi2020,Das2020,Horta2021,Matsuno2022,Horta2023}). 

The smooth, global structure of a stellar halo also reflects how a galaxy formed.
Indeed, the results of numerical simulation for galaxy formation suggest a relationship of 
the smooth component of a stellar halo and the past merging history \citep{Bekki2001,Bullock2005,Deason2014,Pillepich2014,Grand2017,Monachesi2018}.
For instance, \citet{Deason2014} showed from the simulation results of \citet{Bullock2005} that the slope of the density
profile for the outer part of a stellar halo depends on the average time of merging: a more recent merging time reveals a
shallower radial density profile at the Galactocentric distance, $r$, beyond 50 kpc. Also, the slope of a stellar halo profile
may show a break, and it can be a signature of the past merging event, where a merging progenitor stellar system leaves
its debris at its apocenter position \citep{Deason2018b}. 

More recent suite of hydrodynamical simulations for galaxy formation by Rodriguez-Gomez et al. (2016) using the Illustris
Project shows that the so-called {\it in situ} halo, i.e., main progenitor halo, shows a steep density
profile below a transition radius (roughly $r \sim 50$ kpc for the MW-sized halo), whereas the {\it ex situ} halo beyond
this radius mainly originating from the accretion of small stellar systems exhibits a shallow slope having an outer boundary
(see also \cite{Rey2022} for more details on the dependence on merging histories).
Also, based on the magneto-hydrodynamical numerical simulation for galaxy formation named Auriga,
\citet{Grand2017} and \citet{Monachesi2018} showed that both the slope in a density profile of a simulated stellar halo and
its metallicity gradient are intimately related to the number of main progenitor satellites, which contribute to
the total mass of a final halo.

In addition, a sharp jump in the density profile named splashback radius, which is related to the edge of the halo, 
also plays an important role in elucidating the formation history and mass of galaxies, where stars and/or dark-matter particles
reach the apocenter of their first orbit: this physical boundary separates the infalling/accreting material from the halo.
For dark matter, \citet{Adhikari2014}, \citet{Diemer2014} and \citet{More2015} explored the density profile and found that
the splashback radius falls in the range $0.8-1.0$ $r_{200}$ for rapidly accreting halos, and is $\sim 1.5$ $r_{200}$
for slowly accreting halos, where $r_{200}$ denotes the virial radius inside which the mean density is 200 times that of
the Universe. \citet{Deason2020} mentioned that in the case of an environment like the Local Group,
the second jump in the density profile of the star is shown to be more observable.
It is thus of great importance to derive the global smooth structure of a stellar halo to infer its merging history.

For this purpose, the stellar halo in the MW is an ideal target, because its three-dimensional distribution is available
from several halo tracers, such as red giant-branch stars (RGBs), RR Lyrae (RRLs), blue horizontal-branch stars (BHBs) and
blue straggler stars (BSs). These stars are bright enough to map out the stellar halo in the MW up to its outermost edge 
(e.g., \cite{Sluis1998,Yanny2000,Chen2001,Sirko2004,Newberg2005,Juric2008,Keller2008,
Sesar2011,Deason2011,Xue2011,Deason2014,Cohen2015,Cohen2017,Vivas2016,Slater2016,
Xu2018,Hernitschek2018,Iorio2018,Fukushima2018,Fukushima2019,Stringer2021,Yu2024,Medina2024,Feng2024,Amarante2024}). 
From these studies, the smooth component of the stellar halo can be fitted to a power-law density profile with $r^{-\alpha}$,
where $\alpha$ is a power-law index, after the removal of several halo substructures including the Sagittarius stream
and the Virgo overdensity \citep{Ibata1995,Belokurov2006,Juric2008}.

Recent large grand-based surveys for BHBs such as Canada-France Imaging Survey (CFIS), Dark Energy Survey (DES), and
Subaru/Hyper Suprime-Cam Strategic Program (HSC-SSP) have succeeded to map out the outer regions of the stellar halo
beyond $r \sim 40$~kpc \citep{Hernitschek2018,Deason2018a,Fukushima2018,Thomas2018,Fukushima2019,Yu2024}.
For instance, \citet{Thomas2018} selected BHBs from the CFIS $u$-band imaging data combined with $griz$-band data
from Pan-STARRS~1 and showed that the stellar halo represented by a broken power-law profile with an inner/outer
slope of $4.24/3.21$ at a break radius of 41.4 kpc is the best fitting case out to $r \sim 220$~kpc.
More recently, \citet{Yu2024} using DES DR2 $griz$ data selected $\sim 2,100$ BHBs and derived $\alpha$ of $\sim 4.28$
over $20 \lesssim r \lesssim 70$~kpc.
The HSC-SSP data were also used for the selection and analysis of BHBs from the $griz$ imaging data 
\citep{Deason2018a,Fukushima2018,Fukushima2019} to derive the stellar halo up to $r \sim 300$~kpc.
\citet{Fukushima2019} reported that the smooth stellar halo beyond $r \sim 50$~kpc can be represented as 
a single power-law model with $\alpha \simeq 3.7$  and an axial ratio of $q \simeq 1.7$ or alternatively
a broken power-law model with an inner/outer slope of $2.9/15.0$ at a break radius of $160$kpc, based on 
the internal data release of HSC-SSP obtained before 2018 April, which covered over the sky coverage of $\sim 550$~deg$^2$.
This work reported here is an extension of our previous work in \citet{Fukushima2019} using the latest (being almost the final)
internal release of the HSC-SSP data over the sky area of $\sim 1,100$~deg$^2$, i.e., twice as large as the area used
in \citet{Fukushima2019}, to arrive at the most likely density profile of BHBs in the outermost parts of the halo.

This paper is organized as follows.
Section 2 explains the HSC photometry data that we adopt here and the method for the selection
of candidate BHBs based on the $griz$-band photometric data from the HSC-SSP survey.
In this section, we also describe our Bayesian method for the selection of BHB stars and the derivation of
their spatial distribution. In Section 3, we show the results of our Bayesian analysis
for the best set of parameters of the spatial distribution of BHB stars.
In Section 4, the discussion of the current results is presented, and our conclusions are drawn in Section 5.

\begin{table*}
\tbl{Obseved Regions with HSC-SSP}{%
\begin{tabular}{lcccccc}
\hline
Region &   RA  &  DEC  & Adopted area \\ 
       & (deg) & (deg) &  (deg$^2$)   \\
\hline \hline
Spring field   &   $130-230$ & $(-2)-5$   & $\sim 699$  \\
Autumn field   &  $330-40$   & $(-7.5)-6$ & $\sim 559$  \\
HECTOMAP  &  $200-250$  &   $42-45$    & $\sim 109$ \\
\hline
AEGIS     &  $213-216$ &   $51.7-53.3$ & 2.5 & \\
\hline
\end{tabular}}\label{tab: region}
\end{table*}

\section{Data and Method}

\subsection{Data}

We adopt the imaging data obtained from the HSC-SSP Wide layer covering
$\sim 1,100$~deg$^2$ in five photometric bands ($g$, $r$, $i$, $z$, and $y$)
\citep{Aihara2018a,Aihara2018b,Furusawa2018,Kawanomoto2018,Komiyama2018,Miyazaki2018}, where
the target 5$\sigma$ point-source limiting magnitudes are ($g$, $r$, $i$, $z$, $y$) = (26.5, 26.1, 25.9, 25.1, 24.4) mag.
In this work, we utilize the $g$, $r$, $i$ and $z$-band data released in 2021 January
(internal data release S21A), for the selection of most likely BHBs and the removal of
other contaminants as explained below.

\begin{figure}
    \begin{center}
    \includegraphics[width=0.9\linewidth]{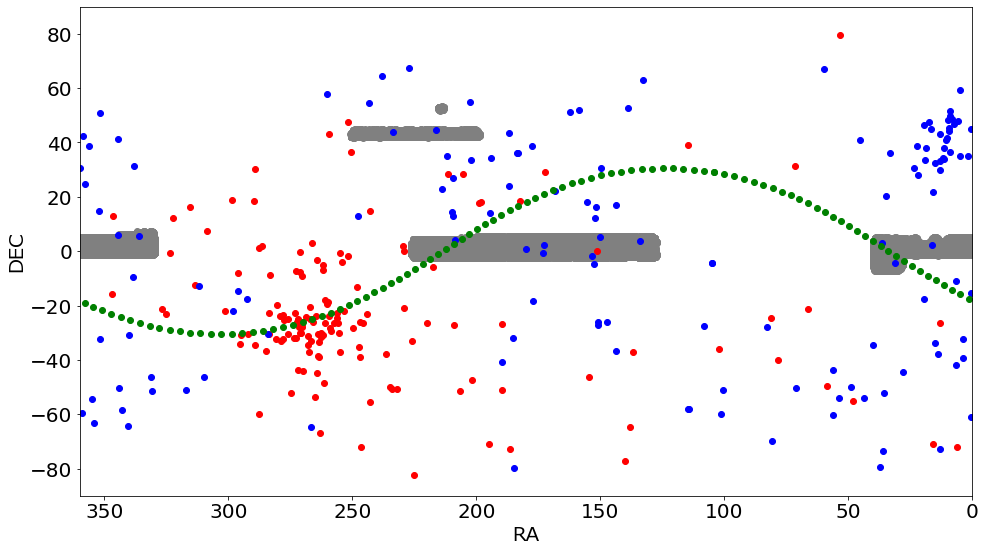}
    \includegraphics[width=0.9\linewidth]{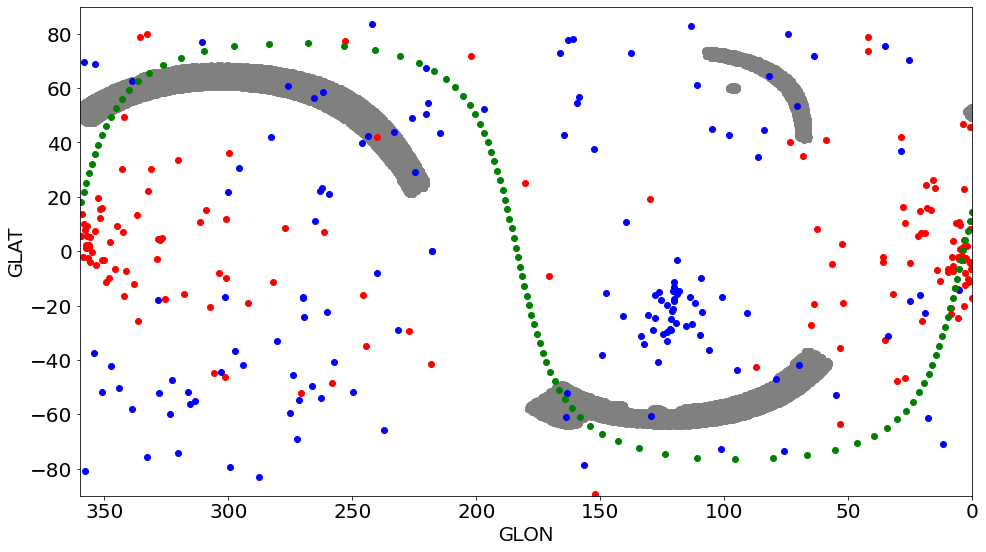}
    \end{center}
    
    \caption{ Upper and middle panels show observed areas of HSC-SSP S21A Wide layer (black shaded regions), 
    compared with the distributions of globular 
    clusters (red dots: \citet{Harris1996})
    and dwarf galaxies (blue dots: \citet{McConnachie2012}). 
    The green dots trace the Sagittarius stream
    (\citet{Belokurov2014}).

    }
    \label{fig:region}
\end{figure}

In S21A, the data set covers mainly two separate wide fields (Figure~\ref{fig:region}), i.e., the north and south sides of the Galaxy.
The former and latter are named Spring field at
$(\alpha_{\rm 2000},\delta_{\rm 2000})=(130^{\circ}-230^{\circ},(-2^{\circ})-5^{\circ})$,
and Autumn field at
$(\alpha_{\rm 2000},\delta_{\rm 2000})=(330^{\circ}-40^{\circ},(-7.5^{\circ})-6^{\circ})$, respectively (see Table~\ref{tab: region}).
The HSC data are processed with hscPipe 8.5.3 \citep{Bosch2018}, a branch of the Large Synoptic Survey Telescope pipeline
\citep{Ivezic2008,Axelrod2010,Juric2017,Ivezic2019} calibrated against PS1 DR1
photometry and astrometry \citep{Schlafly2012,Tonry2012,Magnier2013}.
All the photometry data are corrected for the mean Galactic foreground extinction \citep{Schlafly2011}.

In both Spring and Autumn fields, 
there exist several spatial substructures associated with the Sagittarius (Sgr) stream \citep{Belokurov2014},
which is formed from a tidally disrupting, polar-orbit satellite, Sgr dwarf.
Since our aim in this paper is to deduce the structure of the smooth halo component,
we exclude the areas including these substructures in the following analysis 
in a similar way to \citet{Fukushima2018} and \citet{Fukushima2019}.

As shown in Figure \ref{fig:region}, the areas, where globular clusters \citep{Harris1996}, dwarf galaxies \citep{McConnachie2012}, 
and stellar streams have been identified, are excluded from the data to be analyzed.

\subsection{Selection of targets}

As shown in \citet{Fukushima2019}, we select point sources
using the {\it extendedness} parameter from the pipeline, namely
{\it extendedness}$=0$ for point sources and {\it extendedness}$=1$ for
extended images like galaxies.
This parameter is computed based on
the ratio between PSF and cmodel fluxes \citep{Abazajian2004}, where a point source
is defined to be an object having this ratio larger than 0.985. As shown in \citet{Aihara2018b},
this star/galaxy classification becomes uncertain for faint sources.
Therefore, we probabilistically handled the influence of galactic 
contamination using the function shown in Fig.1 of \citet{Fukushima2019}.

We then select point sources in the following magnitude and color ranges:
\begin{eqnarray}
 18.5 < &g& < 24.5 \nonumber \\
 -0.3 < g&-&r < 0 \nonumber \\
-0.4< r&-&i <0.4 \nonumber \\ 
 -0.25< i&-&z <0.1  \ ,
\label{eq: sampleslection}
\end{eqnarray}
where the faint limit for the $g$-band magnitude range is taken based on its photometric error of
typically $\simeq 0.05$ mag with maximum of $\simeq 0.1$ mag.

These point sources include not only BHBs but also other point sources including BSs,
WDs and QSOs, with some amount of faint galaxies which are missclassified as stars.
As demonstrated in \citet{Fukushima2018}, BHBs are distributed in the distinct region in the
$i-z$ vs. $g-r$ diagram, because the $i-z$ color is affected by the Paschen features of stellar spectra
and is sensitive to surface gravity \citep{Lenz1998,Vickers2012}. Thus, other A-type stars having
higher surface gravity, i.e. BSs, as well as white dwarfs (WDs) can be excluded based on their distributions
in the $i-z$ vs. $g-r$ diagram. Since quasars (QSOs) are largely overlapping with BHBs in this diagram,
the removal of these point sources also requires the use of the
$r-i$ vs. $i-z$
diagram.

\subsection{Probability distributions of BHBs, BSs, WDs, QSOs and galaxies in the color-color diagrams}

This paper adopts a Bayesian method for the selection of BHB stars as shown in \citet{Fukushima2019}.
In this method, the likely distributions for each of BHBs, BSs, WDs, QSOs and faint galaxies in the color-color diagrams
defined by $g$, $r$, $i$ and $z$-band are assembled and analyzed.

For this purpose, we perform a crossmatching between the HSC-SSP data and the corresponding datasets for these objects
taken from several other works, i.e., WDs taken from \citet{Kleinman2013,Kepler2015,Kepler2016},
QSOs obtained in \citet{Paris2018}\footnote{\url{http://www.sdss.org/dr14/algorithms/qso_catalog}}.
After this crossmatching with HSC-SSP, we obtain 401 WDs and 3705 QSOs.

To get probability distributions of BHBs and BSs, we use the same samples as in \citet{Fukushima2019},
which utilized the data provided from SEGUE (Sloan Extension for Galactic
Understanding and Exploration) Stellar Parameter Pipeline (SSPP: \cite{Lee2008}).

We then set the constraints of $3.0 < \log (g) < 3.75$ for BHBs and $3.75 < \log (g) < 4.5$ for BSs,
which well separate the both stellar populations.
The constraints for this selection are 
similar to
those in \citet{Vickers2012},
which set $3.0< \log (g) <3.75$ for BHBs and $3.75 < \log (g) < 5.0$ for BSs.

\begin{figure*}[t!]
\begin{center}
\includegraphics[width=170mm]{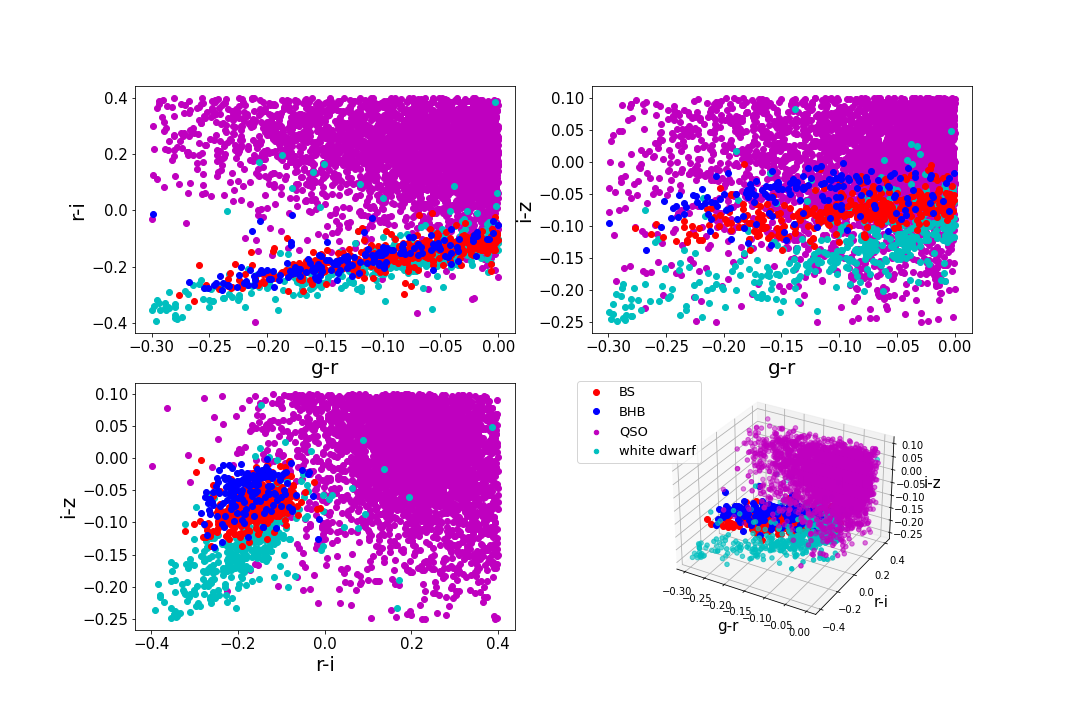}
\end{center}
\hspace*{10mm}
\caption{
The color-color diagrams for each of objects, WDs (cyan), QSOs (magenta), BSs (red) and
BHBs (blue circles) in the $g-r$ vs. $r-i$ space (upper left panel), the $g-r$ vs. $i-z$ space
(upper right panel) and the $r-i$ vs. $i-z$ space (lower left panel).
The lower right panel shows the three dimensional diagram in the $g-r$, $r-i$ and $i-z$ colors.
It follows that we can distinguish these objects in these color-color diagrams.
}
\label{fig: color-color}
\end{figure*}

For galaxies as remaining contaminants, we use the HSC-SSP data with {\it extendedness}$=1$,
corresponding to extended images. 

Figure \ref{fig: color-color} shows the locations of BHBs, BSs, WDs and QSOs
in the color-color diagrams defined with $g$, $r$, $i$ and $z$-band. 
It follows that we can separate QSOs from other objects using $r-i$ color
and classify BHBs, BSs and WDs using $i-z$ color, as mentioned in the previous subsection. 

Next, we use these distributions of different objects in the color-color diagrams
for the application of a Bayesian method described in \citet{Fukushima2019}.
We construct the probability distribution function, $p(griz \mid {\rm Comp})$, 
for each population(${\rm Comp}=$QSO, WD, BHB, BS and galaxy)
in terms of the mixture of several Gaussian distributions.
For this purpose. we use an extreme deconvolution Gaussian mixture model
(XDGMM\footnote[2]{\url{https://github.com/tholoien/XDGMM}}; 
\citet{Bovy2011} and \citet{Holoien2017}) with Python module.

Also to reduce contamination effect of galaxies, we use the same function and fitting parameters for
the fraction of stars classified as HST/ACS among HSC-classified stars
and we fit this fraction with the following function:
\begin{eqnarray}
P_{\rm star}(i) = \frac{1}{1 + \exp(ai + b)} \ ,  
\label{eq: Pstar}
\end{eqnarray}
where $i$ represent $i$-band magnitude and $(a, b)$ are the free parameters.

\subsection{Distance estimates and spatial distributions for sample objects}

In addition to the probability distribution in the color-color diagrams,
we require the density distribution
for each population as functions of the $g$-band magnitude and spatial coordinates.

For both QSOs and galaxies, we assume, for simplicity, a constant density distribution
without depending on the $g$-band magnitude and spatial coordinates, although there may
exist some large scale structures.

For WDs, we adopt a disk-like spatial distribution given by \citet{Juric2008}, as also used by
\citet{Deason2014}, which assumes an exponential profile and has contributions
from thin and thick disk populations. Using the cylindrical coordinates $(R,z)$,
\begin{eqnarray}
\rho_{\rm thin} &=& \exp (R_0/L_1) \exp (-R/L_1-|z + z_0|/H_1) \nonumber \\
\rho_{\rm thick}&=& \exp (R_0/L_2) \exp (-R/L_2-|z + z_0|/H_2) \nonumber \\
\rho_{\rm disk} &=& \rho_{\rm thin} + \rho_{\rm thick} \ ,  
\label{eq: density_WD}
\end{eqnarray}
where $H_1=0.3$~kpc, $L_1=2.6$~kpc, $H_2=0.9$~kpc, $L_2=3.6$~kpc, $z_0=0.025$~kpc, $R_0=8.5$~kpc.
An absolute magnitude for WDs is taken from the model made by \citet{Deason2014}
with $\log (g_s) =8.0(7.5)$:
\begin{eqnarray}
M_g^{\rm WD} = 12.249 + 5.101(g-r),
\label{eq: Mg_WD}
\end{eqnarray}
where the error is given as $\sigma_{M_g^{\rm WD}} \simeq 0.5$ mag.

For the density distributions of BHBs and BSs,
we assume several models and estimate the associated parameters.
Considering that an axially symmetric broken power-law was the best function in \citet{Fukushima2019},
the functions used for this fitting are shown below;

\begin{itemize}
\item Axially symmetric single power-law (ASPL)
\begin{eqnarray}
\rho_{{\rm halo}}(r_q) \propto r_q^{-\alpha},
   \begin{array}{r} r_q^2 = x^2+y^2+z^2q^{-2} \ , \end{array}
\end{eqnarray}
where $q$ denotes the axis ratio.

\item Axially symmetric broken power-law (ABPL)
\begin{eqnarray}
\rho_{{\rm halo}}(r_q) \propto \left\{ 
\begin{array}{l} r_q^{-\alpha_{{\rm in}}} \ \ \  r_q \leq r_b \\ r_q^{-\alpha_{{\rm out}}} \ \ \  r_q>r_b 
\end{array} \right.
\end{eqnarray}
\end{itemize}

In addition to these models, we also adopt a model with varying flattening, $q(r)$, 
which was utilized in \citet{Hernitschek2018} and \citet{Thomas2018}.
\begin{itemize}
\item Axially symmetric single power-law with varying flattening (ASPL with $q(r)$)
\begin{eqnarray}
&&\rho_{{\rm halo}}(r_q) \propto r_q^{-\alpha} , \\
&&r_q^2 = x^2+y^2+z^2q(r)^{-2} , \\
&&r^2 = x^2+y^2+z^2  , \\
&&q(r) =q_{\infty}-(q_{\infty}- q_0) \exp \left( 1- \frac{\sqrt{r^2+r_0^2}}{r_0} \right)
\end{eqnarray}
where $q_0$ is the flattening at the central region of the halo, 
and $q_{\infty}$ is the flattening at large 
Galactocentric distance. 
$r_0$ is a characteristic radius marking a change between these values.
\end{itemize}

To obtain distance estimates for BHBs, we adopt the formula for their $g$-band absolute magnitudes,
$M_g^{\rm BHB}$, calibrated by \citet{Deason2011},

\begin{eqnarray}
M_g^{\rm BHB} &=& 0.434 - 0.169(g_{\rm SDSS}-r_{\rm SDSS})  \nonumber \\
      & &+ 2.319(g_{\rm SDSS}-r_{\rm SDSS})^2  \nonumber \\
      & &+ 20.449(g_{\rm SDSS}-r_{\rm SDSS})^3  \nonumber \\
     & &+ 94.517(g_{\rm SDSS}-r_{\rm SDSS})^4  ,
\label{eq: Mg_BHB}
\end{eqnarray}

where both $g$ and $r$-band magnitudes are corrected for interstellar absorption.
To estimate the absolute magnitude of BHBs selected from the HSC-SSP data, we use
Equations (\ref{eq: conversion-g}) - (\ref{eq: conversion}) below to translate HSC to SDSS filter system.
We then estimate the heliocentric distances and the three dimensional positions of BHBs
in rectangular coordinates, $(x,y,z)$, for the Milky Way space, where the Sun is
assumed to be at (8.5,0,0)~kpc. 

The main results of our work remain unchanged by adopting the other values of 
of the Galactocentric distance of the Sun within the assumed observational uncertainties
($7.5$ kpc: e.g. \citet{Francis2014}, $8.5$ kpc: e.g. \citet{schonrich2012}).
The vertical position of the Sun with respect to the Galactic disk is also uncertain, 
but it is estimated to be smaller than $50$ pc 
and thus negligible for the purpose of this work \citep{Karim2017,Iorio2018,Hernitschek2018}.

To consider the finite effect of contamination from BS stars
as shown below, we adopt their $g$-band absolute magnitudes, $M_g^{\rm BS}$, given by \citet{Deason2011},
\begin{equation}
M_g^{\rm BS} = 3.108 + 5.495 (g_{\rm SDSS}-r_{\rm SDSS}) .
\label{eq: Mg_BS}
\end{equation}
where the typical error is $\sigma _{M_g^{\rm BS}} \simeq 0.5$.

To estimate their absolute magnitudes, we convert the current HSC filter system to
the SDSS one by the formula given as \citet{Homma2016}
\begin{eqnarray}
g_{\rm HSC} &=& g_{\rm SDSS} - a (g_{\rm SDSS} - r_{\rm SDSS}) - b   \label{eq: conversion-g} \\
r_{\rm HSC} &=& r_{\rm SDSS} - c (r_{\rm SDSS} - i_{\rm SDSS}) - d   \\
i_{\rm HSC} &=& i_{\rm SDSS} - e (r_{\rm SDSS} - i_{\rm SDSS}) + f   \\
z_{\rm HSC} &=& z_{\rm SDSS} + g (i_{\rm SDSS} - z_{\rm SDSS}) - h    ,
\label{eq: conversion}
\end{eqnarray}
where $(a,b,c,d,e,f,g,h) = (0.074, 0.011, 0.004, 0.001, 0.106, 0.003, 0.006, 0.006)$ 
and the subscript HSC and SDSS denote the HSC and SDSS system, respectively.
These formula have been calibrated from both filter curves and spectral atlas
of stars \citep{Gunn1983}.

Defining the likelihood as described in \citet{Fukushima2019},
we estimate parameters for density distributions of BHBs and BSs
using Goodman \& Weare's Affine Invariant Markov chain Monte Carlo (MCMC) \citep{Goodman2010},
which makes use of the Python module emcee\footnote[3]{\url{https://github.com/dfm/emcee}}
\citep{Foreman-Mackey2013} and judge these models based on bayesian information criterion (BIC).
We assume that the prior distribution for model parameters is uniform over the range enough to 
not limit the results (see Table~\ref{tab:prior_distribution}).

\begin{table*}
\tbl{Prior distribution for model parameters}{%
\begin{tabular}{l|l|l|ccc}
\hline
Model &   BHB                    & BS              &
          $f_{\rm BHB}$ & $f_{\rm WD}$ & $f_{\rm QSO}$      \\
\hline\hline
ASPL  &   $\alpha=$2-10, $q=$0.1-4     & $\alpha=$2-10, $q=$0.1-4   &
          0-1           &  0-1         &  0-1               \\
\hline
ABPL  &   $\alpha_{\rm in}=$2-10, $\alpha_{\rm out}=$2-10  &   $\alpha_{\rm in}=$2-10, $\alpha_{\rm out}=$2-10  &
          0-1           &  0-1         &  0-1              \\
      &   $r_{\rm b}/{\rm kpc}=$50-400, $q=$0.1-4          &   $r_{\rm b}/{\rm kpc}=$20-200, $q=$0.1-4          &
                        &              &                    \\
\hline
ASPL &   $\alpha=$2-10, $r_{0}/{\rm kpc}=$50-400  &   
 $\alpha=$2-10, $r_{0}/{\rm kpc}=$20-200  &
          0-1           &  0-1         &  0-1              \\
 with $q(r)$ &   $q_{0}=$0.1-4, $q_{ \infty}=$0.1-4          &   
       $q_{0}=$0.1-4, $q_{ \infty}=$0.1-4           &
                        &              &                    \\
\hline
\color{black}
\end{tabular} }
\label{tab:prior_distribution}
\end{table*}

\begin{table*}
\tbl{Best fit parameters}{%
\begin{tabular}{l|l|l|cccc}
\hline
Model &   BHB                    & BS              &
          $f_{\rm BHB}$ & $f_{\rm WD}$ & $f_{\rm QSO}$ & $\Delta$BIC      \\
\hline\hline
ASPL  &   $\alpha=4.11^{+0.18}_{-0.18}$, $q=1.47^{+0.30}_{-0.33}$
      &   $\alpha=4.19^{+0.09}_{-0.08}$, $q=1.36^{+0.11}_{-0.11}$     &
          $0.192^{+0.021}_{-0.022}$ & $0.879^{+0.004}_{-0.004}$ & $0.279^{+0.005}_{-0.005}$ & 0  \\
\hline
ABPL  &   $\alpha_{\rm in}=3.90^{+0.24}_{-0.3}$, $\alpha_{\rm out}=9.1^{+6.8}_{-3.6}$
      &   $\alpha_{\rm in}=4.17^{+0.10}_{-0.10}$, $\alpha_{\rm out}=11.7^{+5.4}_{-4.8}$  &
          $0.190^{+0.022}_{-0.022}$ & $0.881^{+0.005}_{-0.005}$ & $0.279^{+0.005}_{-0.004}$  & 26  \\
      &   $r_{\rm b}/{\rm kpc}=184^{+118}_{-66}$, $q=1.56^{+0.34}_{-0.23}$  
      &   $r_{\rm b}/{\rm kpc}=128^{+34}_{-28}$,   $q=1.35^{+0.11}_{-0.10}$          &
                        &              &                &    \\
\hline
ASPL &   $\alpha=3.77^{+0.32}_{-0.40}$, $r_{0}/{\rm kpc}=189^{+128}_{-93}$
      &   $\alpha=4.09^{+0.15}_{-0.15}$,  $r_{0}/{\rm kpc}=130^{+45}_{-45}$  &
          $0.193^{+0.002}_{-0.002}$ & $0.880^{+0.006}_{-0.005}$ & $0.279^{+0.005}_{-0.004}$  & 22  \\
with $q(r)$ &   $q_{0}=1.67^{+0.48}_{0.30}$, $q_{ \infty}=0.62^{+0.62}_{-0.40}$  
      &  $q_{0}=1.42^{+0.14}_{0.12}$, $q_{ \infty}=0.62^{+0.62}_{-0.40}$ &
                        &              &                &    \\
\hline
\end{tabular} }
\label{tab:best_fit}
\end{table*}

\section{Results}

In this section, we show our main results following the Bayesian method shown in 
Section 2 and compare with our previous work based on the different method for the 
selection of BHBs using the S21A data of HSC-SSP.
As a result of applying the selection in Equation (\ref{eq: sampleslection}) and 
excluding regions that overlapped with the known structures, 
we finally used $42,420$ stars in our analysis.
As already mentioned, this sample includes BHBs, BSs, WDs, and QSOs.


\subsection{Best fit models for all sample}\label{sec:Best fit}

\begin{figure*}[t!]
\begin{center}
\includegraphics[width=80mm]{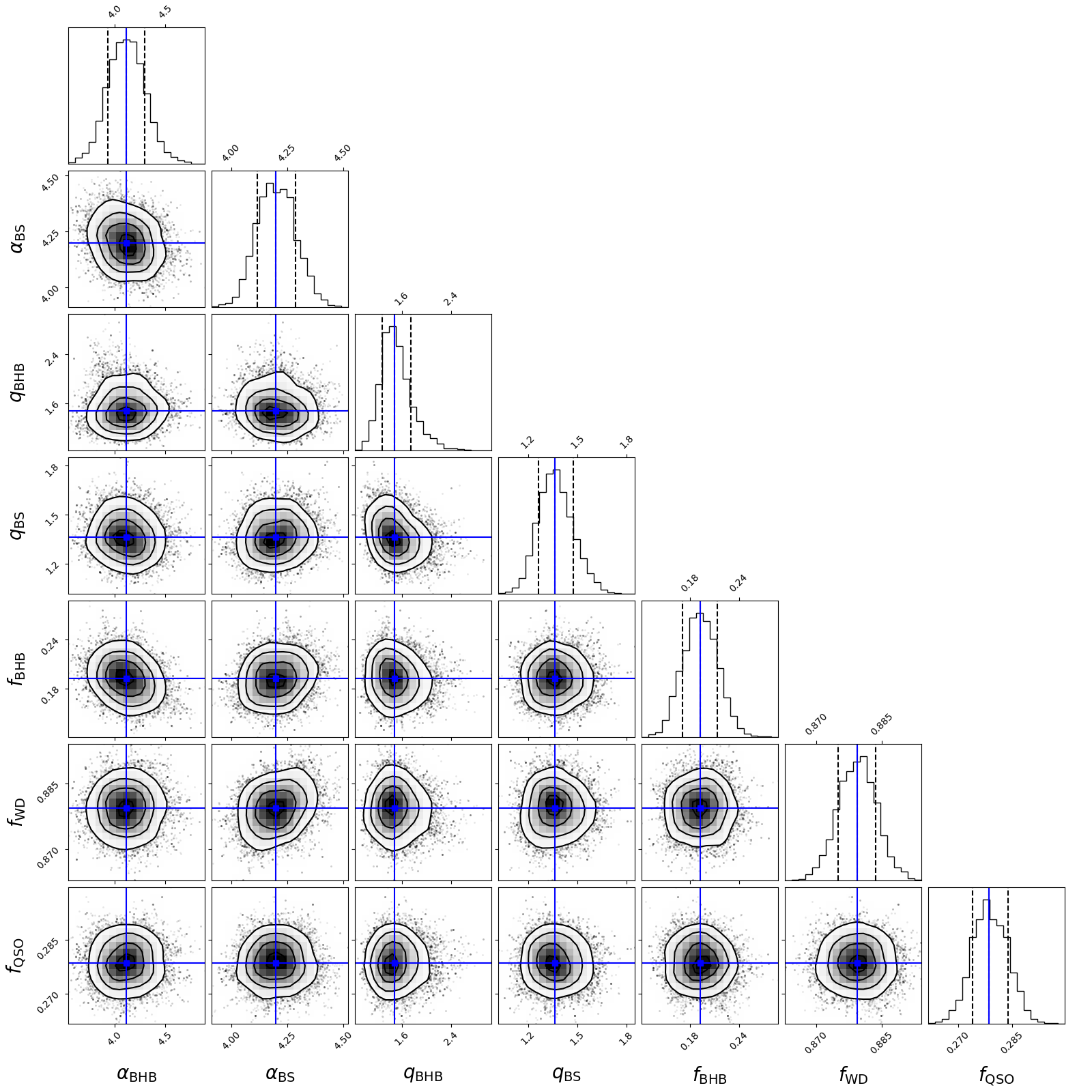}
\includegraphics[width=80mm]{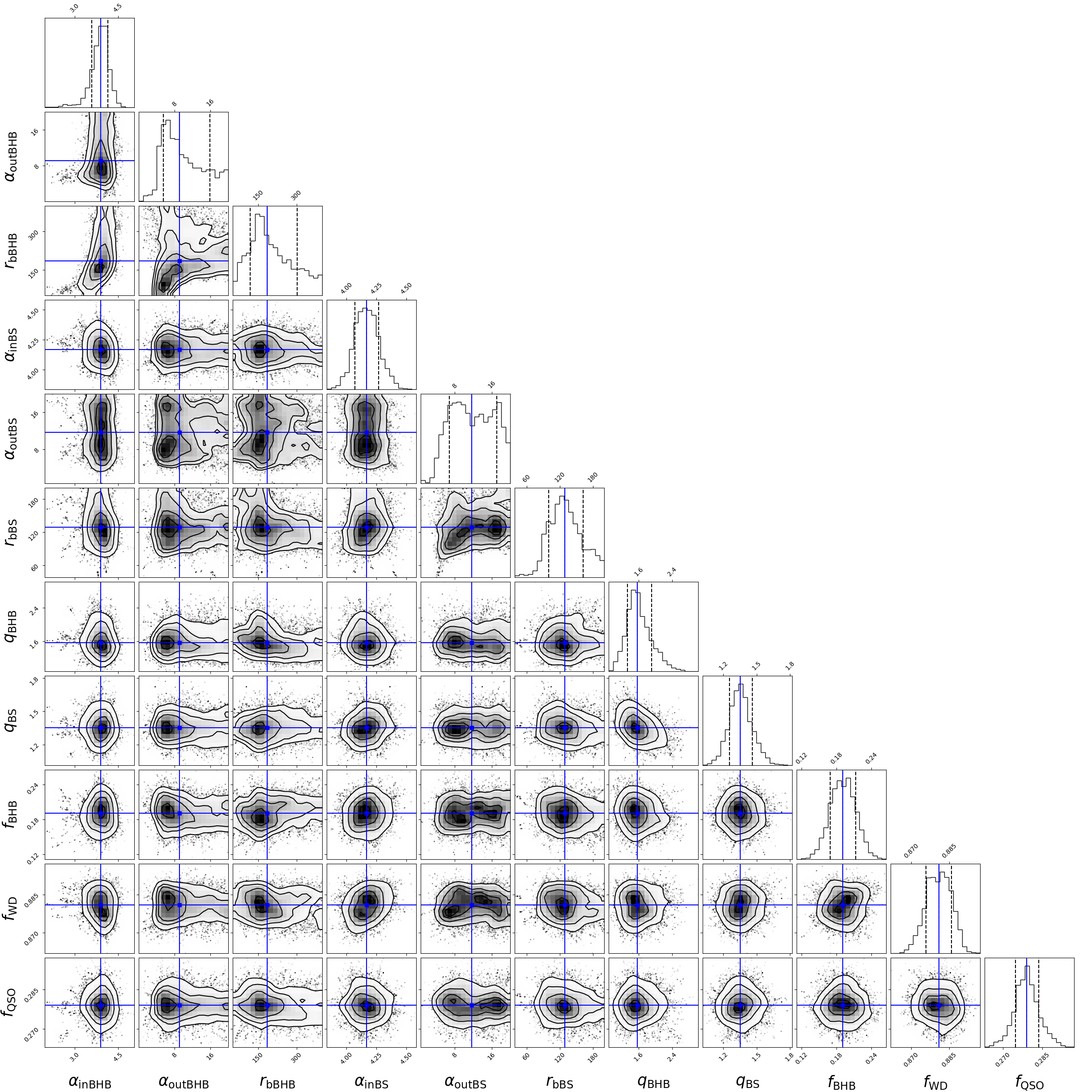}
\end{center}
\begin{center}
\includegraphics[width=80mm]{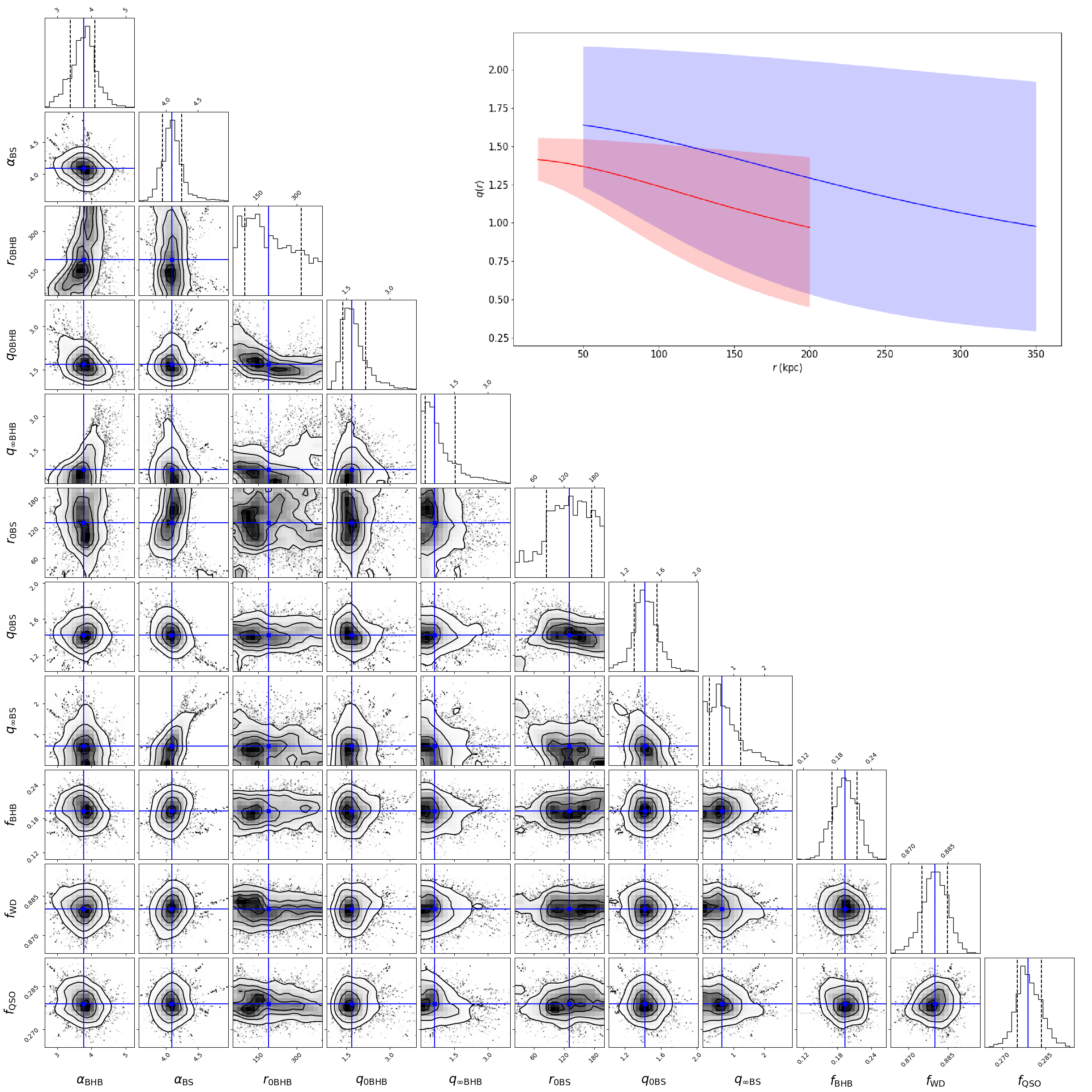}
\end{center}
\hspace*{10mm}
\caption{
MCMC results for ASPL (upper left panel), ABPL (upper right panel) and ASPL with $q(r)$ (lower panel).
}
\label{fig: Axisymmetric}
\end{figure*}

Table \ref{tab:best_fit} shows the best fit parameters for the models of
ASPL and ABPL density profiles, respectively. 
The difference in the BIC values relative to that for the best fit case (ASPL) is also listed in 
the last column and it is defined $\Delta$ BIC. Figures \ref{fig: Axisymmetric} shows the MCMC results for these models. 
We note that as given in Equation (\ref{eq: sampleslection}), these results correspond to the sample with the 
magnitude range of $18.5 < g < 24.5$, suggesting BHBs at about $r = 36 \sim 575$~kpc and 
BSs at about $r = 16 \sim 215$~kpc. The main properties of the results are summarized as follows.
\begin{itemize}
\item Any power-law models reveal similar index values, i.e., 
BHBs are fitted to density profiles with $\alpha = 3.77^{+0.32}_{-0.40} \sim 4.11^{+0.18}_{-0.18}$, whereas BSs show somewhat steeper density profiles
of $\alpha = 4.09^{+0.15}_{-0.15} \sim 4.19^{+0.09}_{-0.08}$.
\item For BHBs, double power-law models (ABPL) show slightly shallower 
profiles at $r < r_{\rm b}$ than the corresponding single power-law models (ASPL)
expressed as $\alpha_{\rm in} < \alpha$. For BSs, $\alpha_{\rm in}$ is 
basically the same as $\alpha$ within the 1$\sigma$ error.
\item Both ASPL and ABPL suggest a prolate shape of $q = 1.47^{+0.30}_{-0.33}$ and $1.56^{+0.34}_{-0.23}$, respectively.
\item ASPL with $q(r)$ also suggests that the MW halo has 
a prolate shape of $q = 1.67^{+0.48}_{0.30}$ at its central region and an oblate shape of $q = 0.62^{+0.62}_{-0.40}$ at large $r$.
\item ASPL provides the lower BIC than ABPL.
\item The best-fit parameters for calculating the fractions of the populations,
$f_{\rm BHB}$, $f_{\rm WD}$ and $f_{\rm QSO}$ are basically the same for different models.
We then obtain the fraction of each population $(\tilde{f}_{\rm BHB},\tilde{f}_{\rm WD},\tilde{f}_{\rm QSO})$,
which are defined in Eqs.(18)-(20) in \citet{Fukushima2019}, as
$\tilde{f}_{\rm BHB}=0.0190-0.0192$,
$\tilde{f}_{\rm WD} =0.0879-0.0881$ and
$\tilde{f}_{\rm QSO}=0.279$.
\end{itemize}

\subsection{Best fit models for north and south parts of the MW}

\begin{figure*}
    \centering
    \includegraphics[width=80mm]{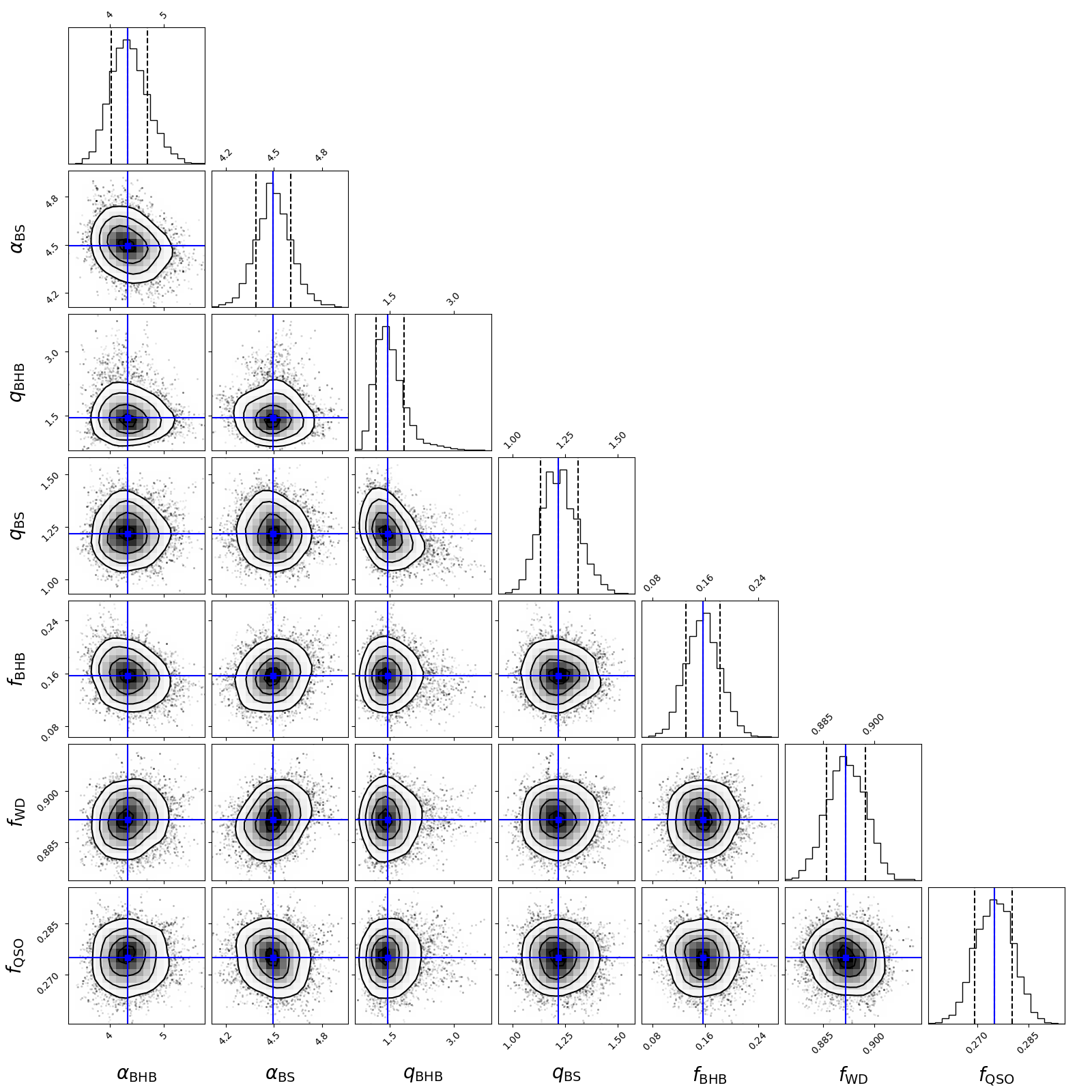}
    \centering
    \includegraphics[width=80mm]{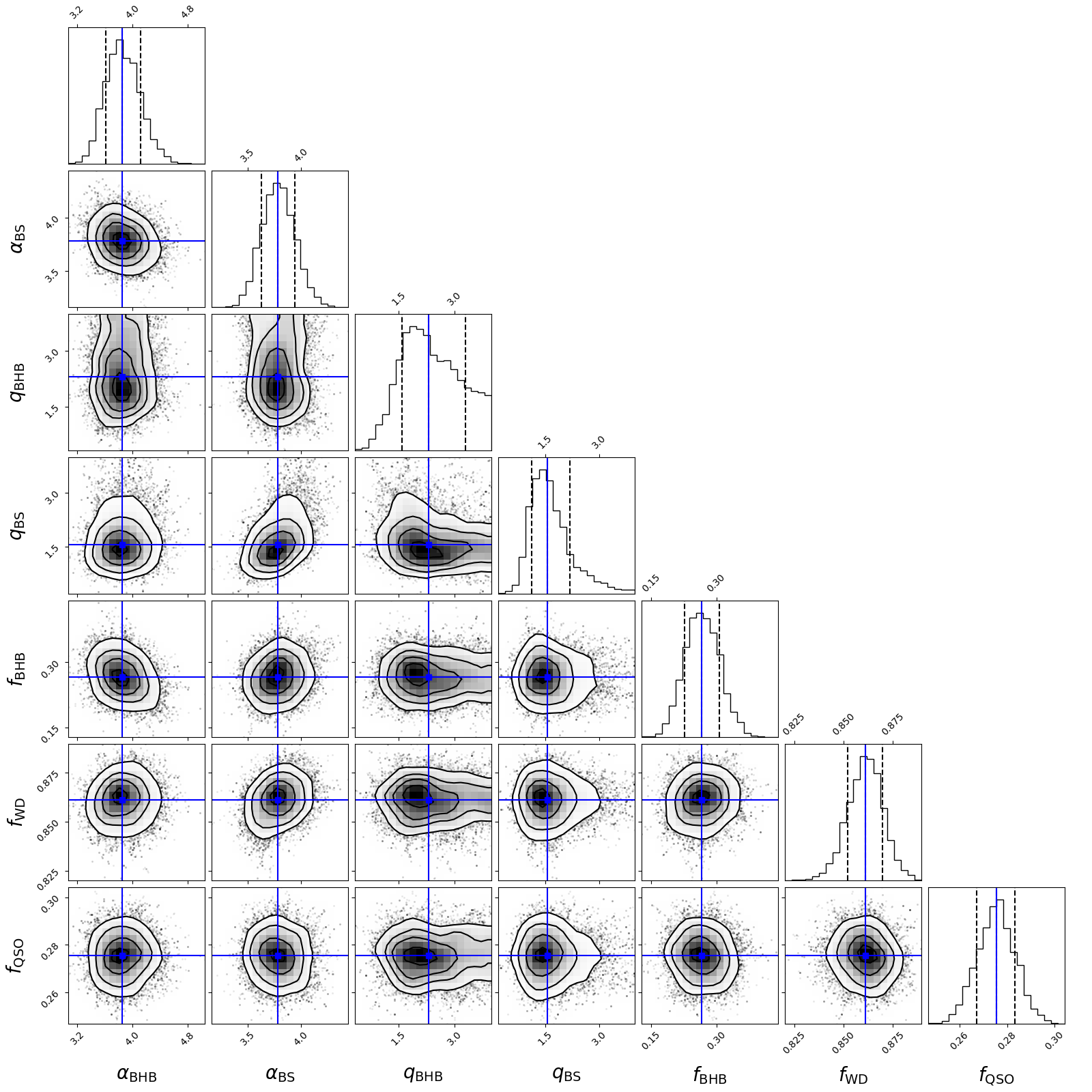}
    \centering
    \includegraphics[width=80mm]{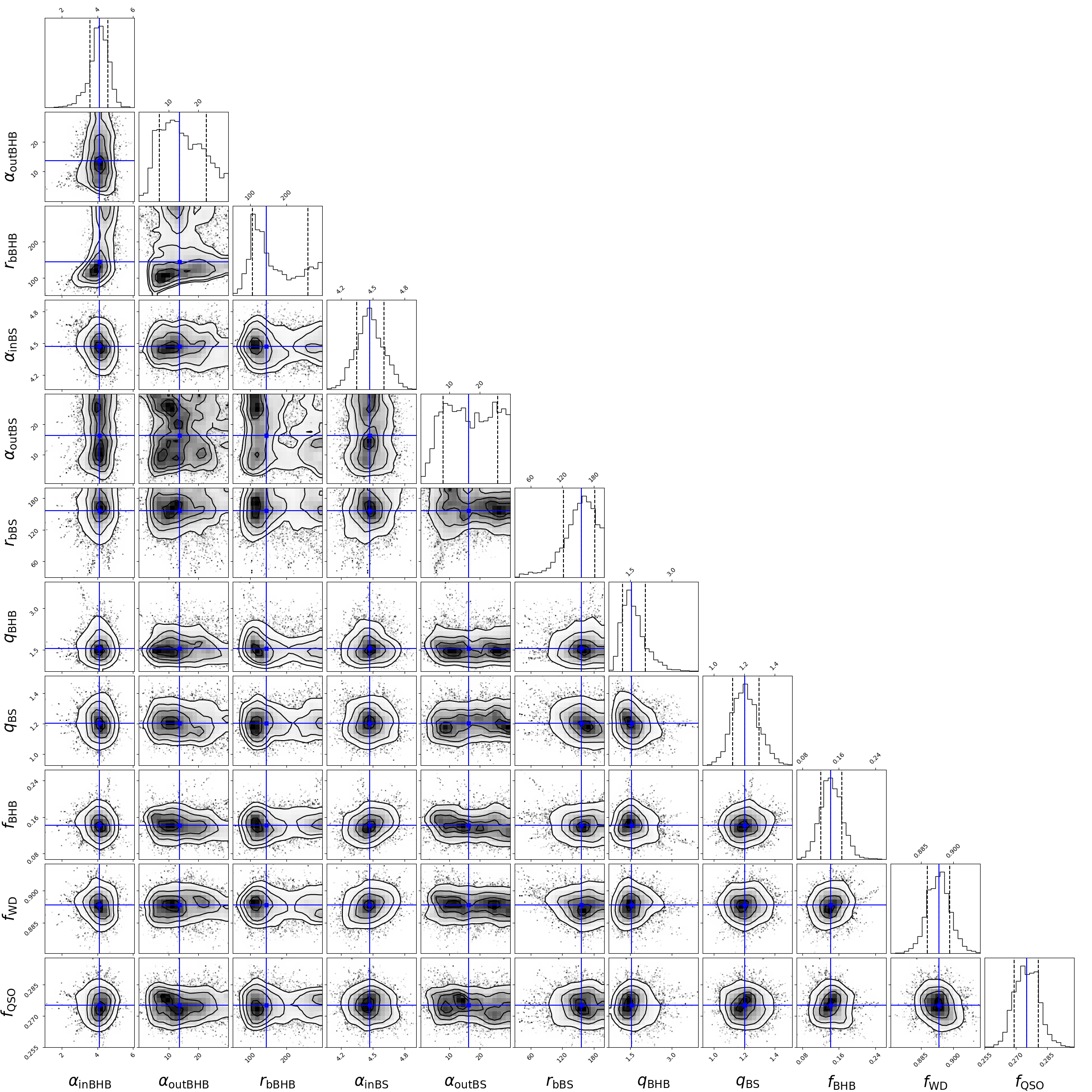}
    \centering
    \includegraphics[width=80mm]{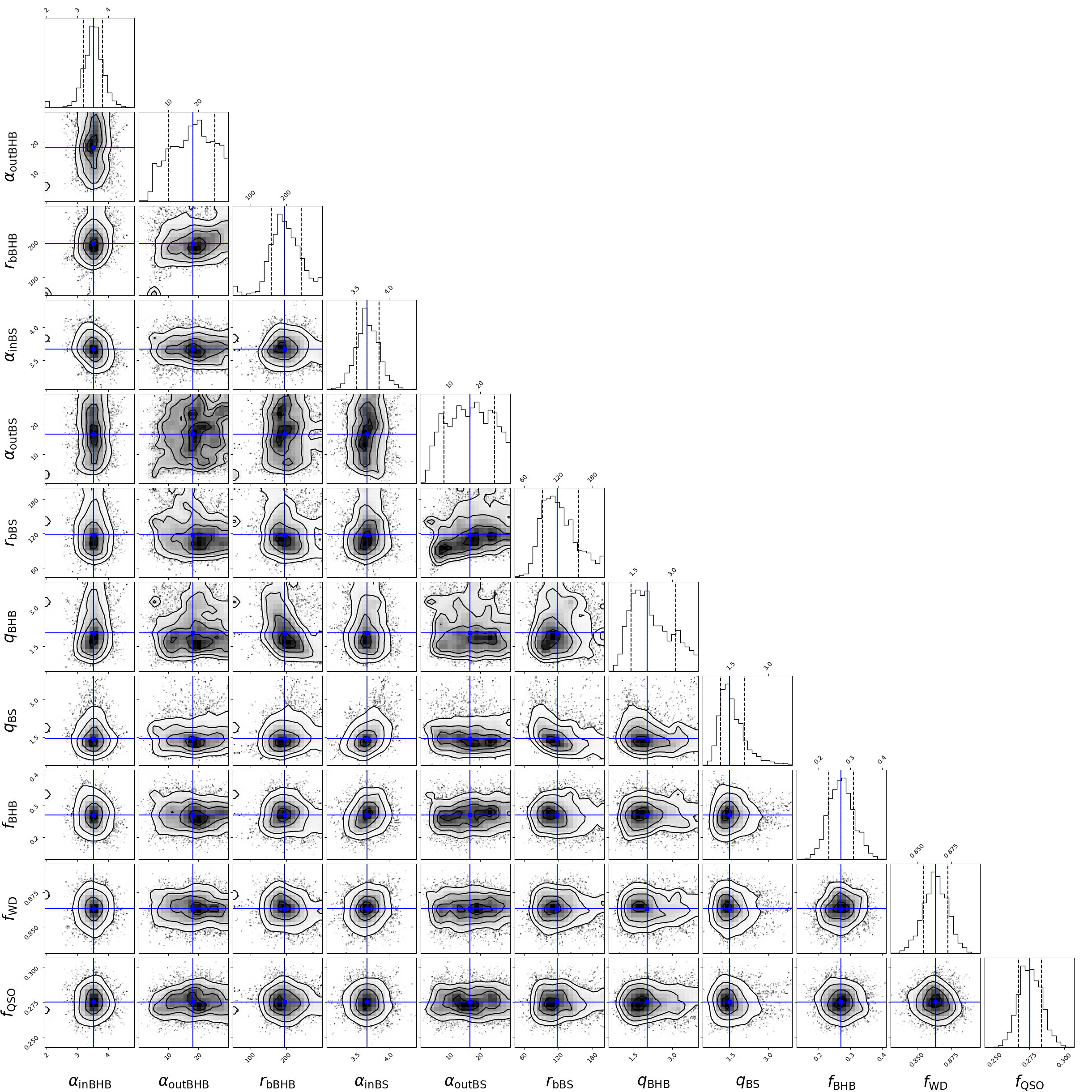}
    \caption{MCMC results for the north (left) and south sides (right panel) of the Milky Way.}
    \hspace*{10mm}
    \label{fig:north-south}
\end{figure*}

We divide the data into the north and south parts of the MW and derive the optimal parameters for 
ASPL and ABPL models of both parts, respectively, to find any asymmetry of the density structure.
For the north part, BHBs are fitted to $\alpha_{\rm in} \simeq 4.11$, 
whereas BSs show a steeper density profile of $\alpha_{\rm in} \simeq 4.47$.
On the other hand, we obtain
$\alpha_{\rm in} \simeq 3.52$ for BHBs,
$\alpha_{\rm in} \simeq 3.67$ for BSs,
in the south part.
The MCMC results for these models are shown in Figures \ref{fig:north-south}. 
The results given above are also tabulated in Table \ref{tab:best_fit_hemisphere} and 
the main results are summarized as follows.

\begin{itemize}
\item The inner regions of both BHBs and BSs are steeper in the north than the south part of the MW.

\item  Both north and south parts show very steep profiles of
$\alpha_{\rm out}$ at distances exceeding $r_{\rm b}= 100 \sim 200{\rm kpc}$.

\item The south part shows $q=1.44 \sim 2.0$, suggesting a quite prolate shape compared to the north part.

\item The south part has a larger fraction of BHBs ($f_{\rm BHB} = 0.26 \sim 0.27$) than the north part
($f_{\rm BHB} = 0.15 \sim 0.16$), suggesting a larger number density of field halo stars in the south part.
\end{itemize}

This signature of asymmetry, especially a more prolate shape and more enhanced fraction of BHBs in the south part,
may be associated with the disturbing effect of the massive Large Magellanic Cloud (e.g. \cite{Garavito-Camargo2019,
Garavito-Camargo2021,Erkal2021}).

\begin{table*}
\tbl{Best fit parameters}{%
\begin{tabular}{l|l|l|l|ccc|c}
\hline
Direction&Model &   BHB                    & BS              &
          $f_{\rm BHB}$ & $f_{\rm WD}$ & $f_{\rm QSO}$ & $\Delta$BIC      \\
\hline\hline
North & ASPL  &   $\alpha=4.34^{+0.36}_{-0.31}$,$q=1.45^{+0.38}_{-0.27}$ 
      &   $\alpha=4.50^{+0.11}_{-0.11}$,$q=1.22^{+0.09}_{-0.09}$   &
          $0.156^{+0.026}_{-0.026}$ & $0.892^{+0.006}_{-0.006}$ & $0.275^{+0.005}_{-0.006}$ & 0 \\
\cline{2-8}
      & ABPL  &   $\alpha_{\rm in}=4.11^{+0.48}_{-0.52}$, $\alpha_{\rm out}=13.6^{+9.0}_{-6.7}$
      &   $\alpha_{\rm in}=4.47^{+0.14}_{-0.12}$, $\alpha_{\rm out}=16.3^{+9.4}_{-8.5}$  &
          $0.142^{+0.025}_{-0.022}$ & $0.893^{+0.005}_{-0.005}$ & $0.275^{+0.006}_{-0.006}$ & 35 \\
      & &   $r_{\rm b}/{\rm kpc}=145^{+115}_{-39}$, $q=1.55^{+0.51}_{-0.33}$  
      &   $r_{\rm b}/{\rm kpc}=156^{+26}_{-34}$, $q=1.20^{+0.09}_{-0.08}$ 
      & & & &\\
\hline
South & ASPL  &   $\alpha=3.85^{+0.27}_{-0.24}$,$q=2.31^{+1.00}_{-0.72}$  
      &   $\alpha=3.79^{+0.16}_{-0.15}$,$q=1.56^{+0.64}_{-0.43}$   &
          $0.267^{+0.041}_{-0.043}$ & $0.861^{+0.008}_{-0.009}$ & $0.275^{+0.008}_{-0.008}$ & 0  \\
\cline{2-8}
      & ABPL  &   $\alpha_{\rm in}=3.52^{+0.29}_{-0.31}$, $\alpha_{\rm out}=18.2^{+7.3}_{-8.2}$
      &   $\alpha_{\rm in}=3.67^{+0.18}_{-0.16}$, $\alpha_{\rm out}=16.7^{+8.1}_{-8.6}$  &
          $0.271^{+0.040}_{-0.039}$ & $0.863^{+0.009}_{-0.009}$ & $0.280^{+0.009}_{-0.008}$ & 8   \\
      & &   $r_{\rm b}/{\rm kpc}=196^{+46}_{-38}$, $q=2.02^{+1.11}_{-0.62}$  
      &   $r_{\rm b}/{\rm kpc}=118^{+38}_{-26}$, $q=1.44^{+0.58}_{-0.36}$ 
      & & & & \\
\hline
\end{tabular} }
\label{tab:best_fit_hemisphere}
\end{table*}

\subsection{Comparison with our previous work}

In our previous paper \citep{Fukushima2019}, we selected candidate BHBs in the range of $18.5 < g < 23.5$
from the S18A data of HSC-SSP over $\sim 550$~deg$^2$ and analyzed them using the same method as adopted in this work.
The results using the S18A data \citep{Fukushima2019} are summarised as 
$\alpha = 3.96^{+0.20}_{-0.16}$ and 
$q = 1.68^{+0.30}_{-0.33}$ for ASPL and 
$\alpha_{\rm in} =2.92^{+0.33}_{-0.33}$, 
$\alpha_{\rm out} = 15.0^{+3.7}_{-4.5}$, 
$q = 1.72^{+0.44}_{-0.28}$ and 
$r_{\rm b} \sim 160^{+18}_{-19}$~kpc for ABPL.
It is interesting to note that $(\alpha, q)$ for ASPL model of the current data (S21A) are basically the same as those of the S18A data,
whereas $\alpha_{\rm in} (= 3.88)$ for ABPL model of the current data is steeper than that of the S18A data and has instead being closer to
$\alpha (= 4.02)$ for ASPL model. This suggests that the global density profile of candidate BHBs can be expressed as $r^{-4}$, having a boundary
at $r_{\rm b} \simeq 200$~kpc with a prolate shape of $q \simeq 1.5-1.7$.
In \citet{Fukushima2019}, ABPL was considered as the best model, 
but in the current work, ASPL is found to be the best model. 
We think that this is due to the difference in the magnitude range, where this work includes fainter stars ($g < 24.5$) than in the previous work ($g < 23.5$).
As proof of this, when we analyze the sample up to the same brightness as the previous work, 
we find that model ABPL is the best model.

\begin{figure*}[t!]
\begin{center}
\includegraphics[width=180mm]{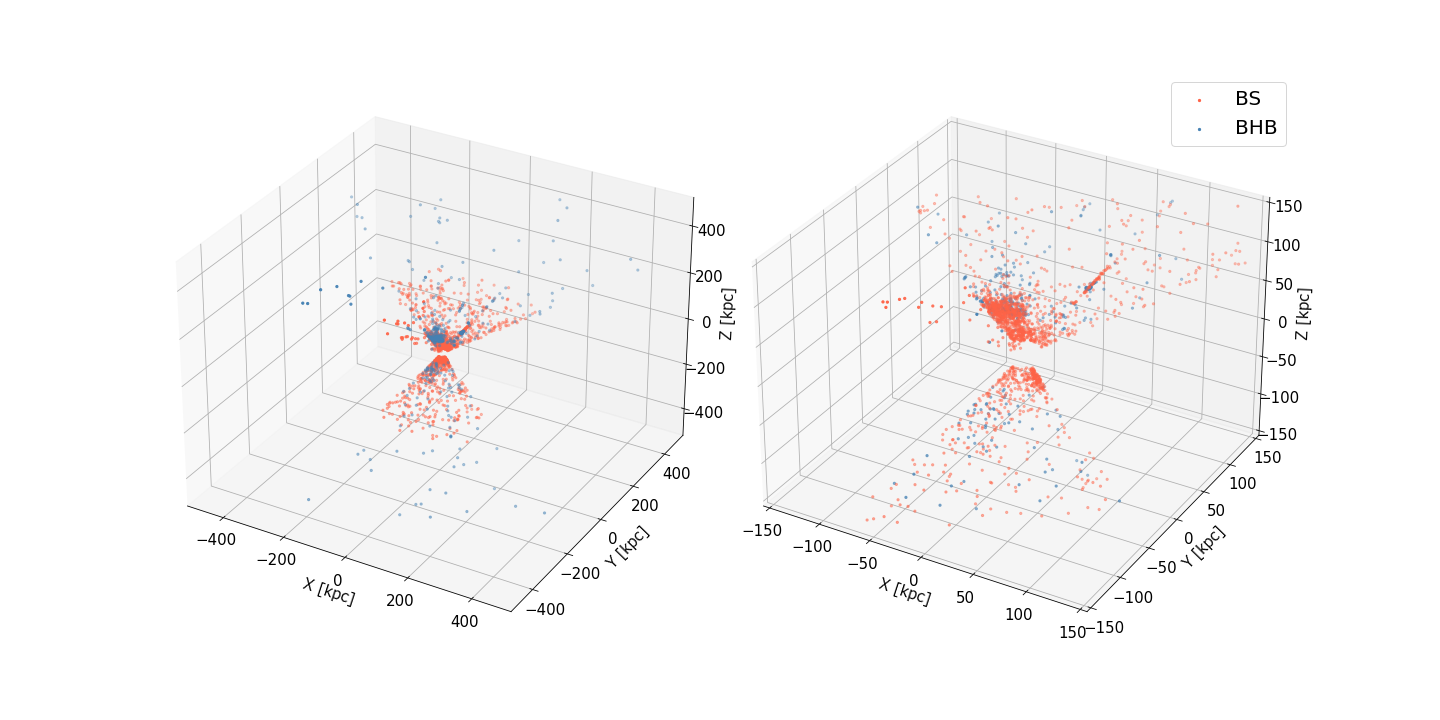}
\end{center}
\hspace*{10mm}
\caption{
Three-dimensional distributions of BHBs (blue points) and BSs (red points) selected from
those having high probabilities as BHBs [$p({\rm BHB}|x)>0.7$] and BSs [$p({\rm BS}|x)>0.7$],
respectively, as defined in Equation (\ref{eq: BHB_prob}).
The left panel shows the box over $-500 \le x,y,z \le 500$~kpc.
The right panel shows the zoom-in view of the inner region over $-150 \le x,y,z \le 150$ kpc.
}
\label{fig: 3d_map}
\end{figure*}


\subsection{Three-dimensional maps of BHBs and BSs}

Given that the parameters $f_{\rm BHB}$, $f_{\rm WD}$ and $f_{\rm QSO}$ basically remain
the same among different density models, it is possible to derive the probability that
a given target is either of a BHB, BS, WD, QSO or galaxy. For instance, the probability
of a BHB is given as
\begin{equation}
p({\rm BHB}|x) = \frac{p(x|{\rm BHB})f_{\rm BHB}}{ \sum_{i=1}^{4}p(x|A_i) \tilde{f}_i 
                                + p(x|{\rm galaxy})\frac{1-P_{\rm star}}{P_{\rm star}} } \ ,
\label{eq: BHB_prob}
\end{equation}
where $x$ shows each sample and $i$ denotes a component (BHB, BS, WD and QSO).

Figure \ref{fig: 3d_map} (with scales of $-500 \le x,y,z \le 500$~kpc and $-150 \le x,y,z \le 150$~kpc
in left and right panels, respectively) show the three-dimensional maps for the sample with $p({\rm BHB}|x)$
larger than 70\% (blue points) and $p({\rm BS}|x)$ larger than 70\% (red points)
using all the survey fields.
As is clear from these figures, there are known substructures (as noted in \cite{Fukushima2019}), the Sgr stream at $(x,y,z) = (-20, 10, 40)$~kpc,
Sextans dSph at $(x,y,z) = (40, 60, 60)$~kpc, and the overdensity at $(x,y,z) = (0, -40, -50)$~kpc.
It is also worth noting that the distribution of candidate BHBs is largely extended beyond $r \sim 150$~kpc as inferred from Figure~\ref{fig: 3d_map}.

\begin{figure*}[t!]
\begin{center}
\includegraphics[width=160mm]{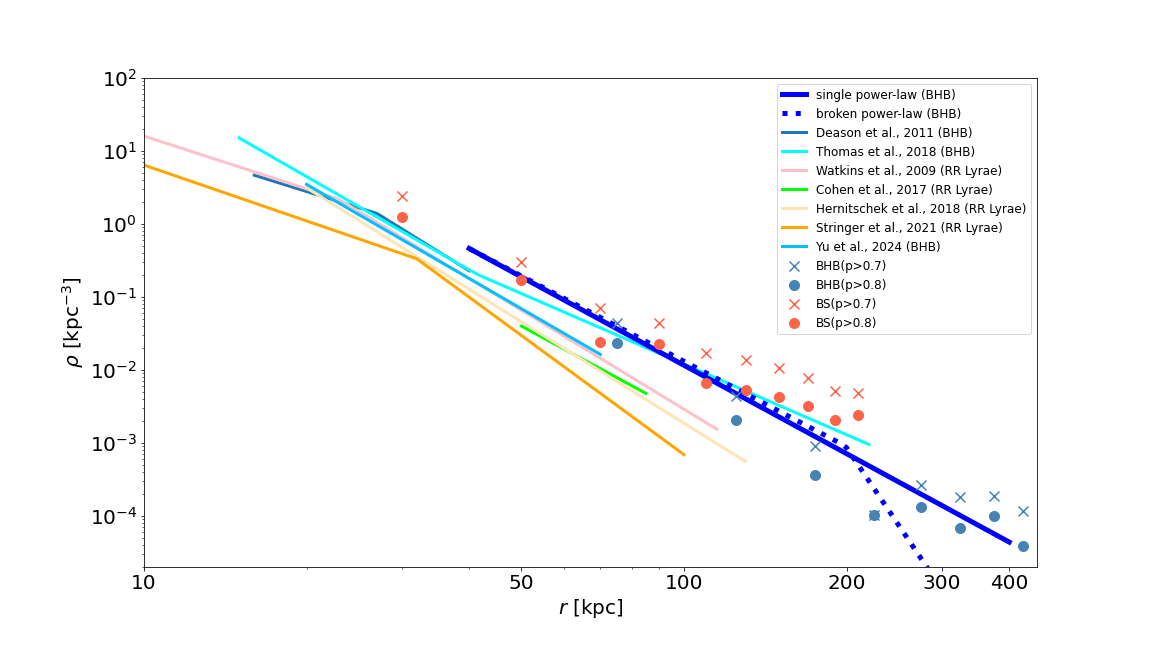}
\end{center}
\hspace*{10mm}
\caption{
Comparison of our best-fit models, the single power-law (blue solid line) and broken power-law
(blue dotted line)
with other works using BHBs \citep{Deason2011,Thomas2018}
and RR Lyrae \citep{Watkins2009,Cohen2017,Hernitschek2018}.
}
\label{fig: comparison}
\end{figure*}

Figure \ref{fig: comparison} shows the radial density profiles of
BHBs (blue dots) and BSs (red dots) obtained in this work, 
where the dots (cross points) correspond to these stars having probabilities
larger than 80\% (70\%), namely $p({\rm BHB}|x)>0.8$ and $p({\rm BS}|x)>0.8$
($p({\rm BHB}|x)>0.7$ and $p({\rm BS}|x)>0.7$).
In 80\% (70\%) of cases, the number of samples selected as BHB are 225 (535) and the number of samples selected as BS are 915 (1,850).
Among these BHBs, the total numbers beyond 100 kpc in all observed fields and the selected fields
(for deriving the global density profile) are 80 (205) and 42 (99), respectively.
For comparison, the survey results of halo tracers in other works are 
also shown in this figure and will be discussed below.  

\section{Discussion}

\subsection{Comparison with other survey results}

As mentioned in Section 1, the structure of the MW stellar halo
has been studied based on the surveys of several halo tracers, as highlighted in Figure \ref{fig: comparison}.

According to recent researches on the same tracer,
\citet{Thomas2018} selected BHBs from their combined dataset 
of the deep $u$-band imaging from the new CFIS and 
the $griz$ bands from Pan-STARRS 1 covering a total of 4,000 ${\rm deg}^2$ of the northern sky and revealed
that a broken power-law model with an inner/outer slope of $4.24/3.21$ at a break radius of 41.4 kpc is
the best fitting case out to $r \sim 220$~kpc.
This outer slope is somewhat shallower than the current result in the corresponding radial interval,
i.e., the inner slope of $\simeq 3.90$
in our ABPL model for all samples. On the other hand, their result is similar to our case when 
the sample is limited to the south part of the MW,
i.e., the inner slope of $\simeq 3.52$ in our ABPL model at $r < r_{\rm b} \simeq 173$~kpc.
Therefore, the results may slightly depend on the sampling area.
In more recent research, \citet{Yu2024} 
revealed power law index $\alpha = 4.28$ 
for stellar halo within a distance range of
$20$ to $70$ kpc using $griz$ photometry data from
the Dark Energy Survey Data Release 2 (DES DR2).

The surveys using RRLs at $r$ as large as 100~kpc tend to provide different density 
slopes \citep{Watkins2009,Cohen2017,Hernitschek2018}. These works show $\alpha = 
4.0 \sim 4.5$ at $r > 25$~kpc, which is systematically steeper than the slopes obtained 
here for BHBs, but consistent with those for BSs located at similar radii to the sample of RRLs 
($\alpha \simeq 4.50$ for ASPL, $\alpha_{\rm out} \simeq 4.22$ for ABPL). 
\citet{Stringer2021} revealed that a broken power-law model with
$\alpha_{\rm in} \simeq 2.54$, $\alpha_{\rm out} \simeq 5.42$, and $r_{\rm b} = 32.1$~kpc is the best fitting case
for RRLs from the full six-year data set of the Dark Energy Survey, which covers 5,000~deg$^2$
of the southern sky out to $r = 335 {\rm kpc}$.
\citet{Medina2024} reported similar results as a break
in the RRLs distribution at 18.1 koc, with an inner slope of $2.05$, and a steeper outer slope of $4.47$,
using Dark Energy Camera data from the High cadence Transient Survey (HiTS) 
and the Halo Outskirts With Variable Stars (HOWVAST) survey.

According to the above studies, the density slope of RRLs is a little steeper than that of BHBs.
This is comparable to the same difference between the results using BHBs and BSs, and this
difference between halo tracers may be due to the process of their formation associated with merging/accretion as discussed below.

It is also worth noting that even beyond the broken radius, there exist some candidate BHBs in the outskirts, whereby
producing a trough around $r \sim 200$ kpc in the density profile. While the origin of these candidate BHBs is not clear
because of their small number, they might be associated with the debris of tidally disrupt dwarf galaxies
or they might be just incompletely remaining contaminants such as distant faint galaxies. To settle this point,
further deep surveys over much larger areas, such as the LSST survey, is needed by increasing the sample of candidate BHBs.

\subsection{Possible constraints on the past accretion history}

The current result for the spatial distribution of BHBs in the outer part of the halo,
characterized by a shallower density profile
than the inner part of the halo ($25 - 50$ kpc), provides an important constraint on the formation history of the MW.
Indeed, numerical simulations for galaxy formation suggest similar halo structures, which is comprised of {\it in situ} and
{\it ex situ} halo components
\citep{Zolotov2009,Zolotov2010,Oser2010,Font2011,McCarthy2012,Tissera2012,Tissera2013,Rodriguez-Gomez2016}. For instance,
\citet{Font2011} and \citet{McCarthy2012} found from their simulated galaxies that the inner halo formed from the {\it in situ}
star formation mode shows a relatively steeper density profile than the outer part of the halo ({\it ex situ} halo), which are
formed from accreted satellites. Similarly, using the Illustris Project \citep{Genel2014,Vogelsberger2014a,Vogelsberger2014b}
for the suite of hydrodynamical simulations for galaxy formation over a wide range of stellar masses,
$M_{\ast}=10^{9}-10^{12} M_{\odot}$, \citet{Rodriguez-Gomez2016} found that these two halo components are spatially separated,
where the {\it in situ} ({\it ex situ}) halo dominates the inner (outer) part of the halo.

Indeed, as shown in Figure 10 of \citet{Rodriguez-Gomez2016} for the total stellar mass of $M_\ast = 10^{11}$ $M_{\odot}$, i.e., like the MW-like stellar
mass, the {\it in situ} halo component is characterized by a relatively shallow density profile over the very inner region of
the halo at $r \lesssim r_{{\rm half},\ast}$ (where $r_{{\rm half},\ast}$ is a stellar half-mass radius) and a steep density
slope over $r_{{\rm half},\ast} \lesssim r \lesssim 4 r_{{\rm half},\ast}$, whereas the {\it ex situ} component at $r \gtsim 4 r_{{\rm half},\ast}$
shows a relatively shallow density slope. This radius ($\sim 4 r_{{\rm half},\ast}$ for $M_\ast = 10^{11}$
$M_{\odot}$) is the transition radius, where the two halo components locally become equally abundant, and this is smaller for
more massive galaxies \citep{Rodriguez-Gomez2016}.
It is worth noting that this behavior of the theoretically predicted halo is generally
in agreement with the observed density profile of the halo probed with RRLs, BSs, and BHBs: as presented in \citet{Deason2014}
(see their Figure 7), the observed halo with BHBs and BSs selected from the Sloan Digital Sky Survey (SDSS) sample shows a shallow profile at
$r \lesssim 25$~kpc ($\alpha \sim 2.5$) and a steep profile over $25 \lesssim r \lesssim 100$~kpc ($\alpha = 4 \sim 6$).
Deeper surveys of these halo tracers like the current work using Subaru suggest a shallow profile beyond $r \sim 100$~kpc
($\alpha \sim 4$), which is possibly the {\it ex situ} component originated from accretion of small stellar systems.

\subsection{The color gradient of BHBs}

\begin{figure*}[t!]
\begin{center}
\includegraphics[width=80mm]{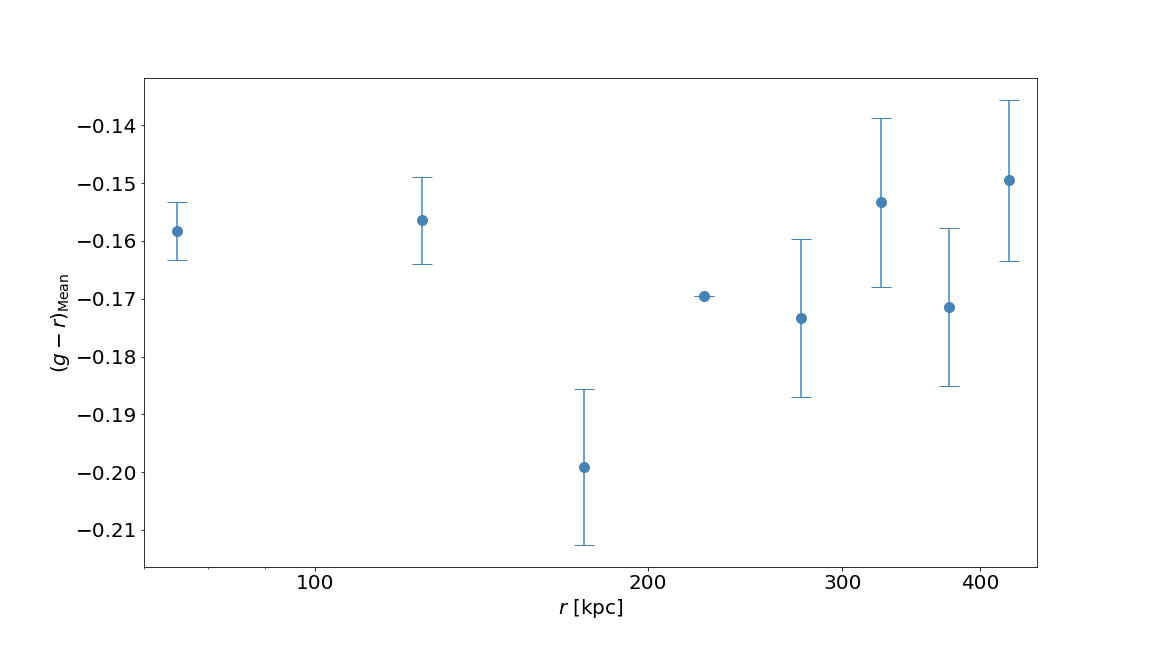}
\includegraphics[width=80mm]{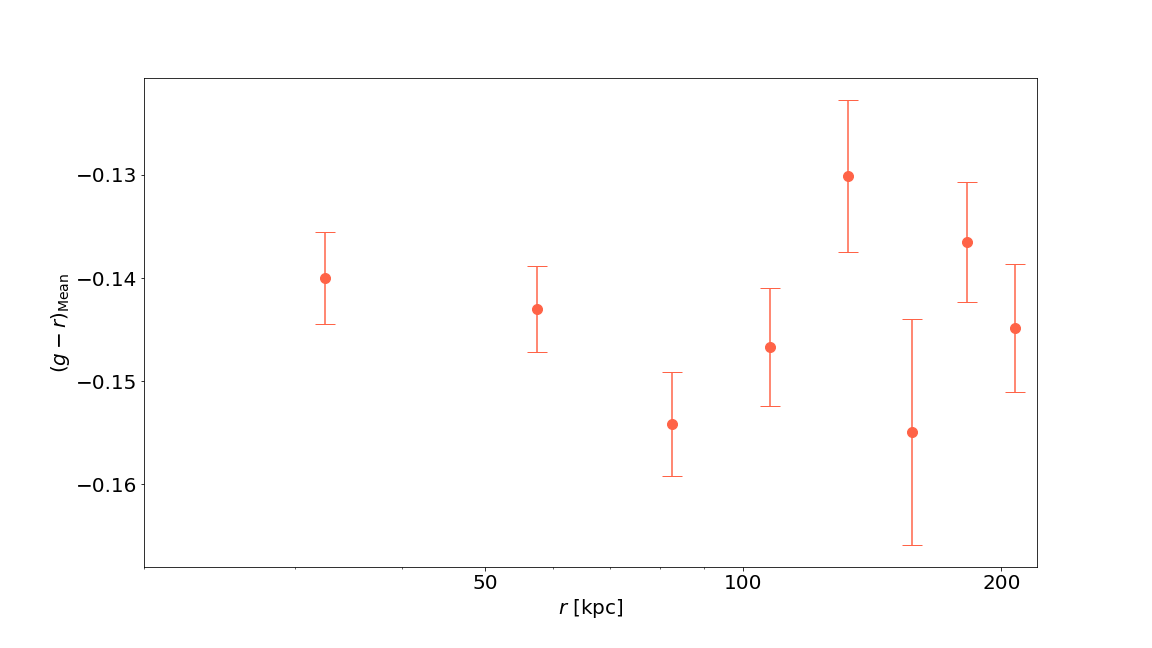}
\end{center}
\hspace*{10mm}
\caption{
The left (right) panel show the color ($g-r$) gradient of BHBs (BSs)
selected by $p({\rm BHB}|x)>0.7$ ($p({\rm BS}|x)>0.7$).
The average value of ($g-r$) for bins separated by 50 kpc (25 kpc)
is plotted as the points and the bars indicate standard errors 
(calculated by $\sigma/\sqrt{n}$ when $\sigma$ is standard 
deviation and $n$ is sample size). We see that rapidly turning 
blue around 180 kpc in the color gradient of BHBs.
}
\label{fig:color_gradient}
\end{figure*}

To infer the information on the spatial distribution of ages and metallicities for BHBs (BSs),
we calculate the average values and standard errors of $g-r$ colors
for bins separated by 50 kpc (25 kpc).
In Figure \ref{fig:color_gradient}, we show the mean color ($g-r$) as a function of $r_q$ for BHBs and BSs
selected by $p({\rm BHB}|x)>0.7$ and $p({\rm BS}|x)>0.7$, respectively ($q = 1.5$ for BHB, 
$q = 1.36$ for BS).
We see that rapidly turning blue around 180 kpc in the color of BHBs.
We adopt a t-test to check whether 
there is a significant difference between the overall 
average of the color and the average in the range of $r=150 \sim 200$ kpc, 
and find that the p-value is $\sim0.018$, indicating it is statistically significant.

In previous studies, the color distribution of BHBs further inside the MW have been analyzed.
For example, \citet{Preston1991} showed a color gradient out to $\sim 12$ kpc.
Furthermore, similar results are obtained $\sim 40$ kpc in \citet{Santucci2015}
using BHB stars selected from the SDSS sample.
They detected that  the mean de-reddened $g-r$ color increases outward in the Galaxy.
In these studies, the trends in color gradients did not differ between samples of
different metallicity, suggesting that the origin of color gradients may 
reflect age gradients.
These may reflect that more massive systems,
which would have  grown over a longer timescale,
are less affected by tidal forces and can fall deeper into the potential.
A similar relationship between age and distance is also shown in \citet{Das2016}
using BHBs from the Sloan Extension for Galactic Understanding 
and Exploration-II (SEGUE-II) survey.
More recent studies have shown similar color and age gradients in \citet{Whitten2019}
using the SDSS, Pan-STARRS1 and the cross-matched GALEX GUVCat with Pan-STARRS DR1.
When compared, the color near the farthest $30$ kpc is $0.14$ to $0.15$, 
which is close to the value of $0.16$ of the innermost sample in our result.

\subsection{On the edge of the MW's stellar halo}\label{sec:MW edge}

In the cosmological $N$-body simulations of $\Lambda$CDM models for structure formation in the Universe,
a sharp decrease in the slope of the radial density profile is seen at the edge of the virialized dark halo,
which usually appears at $r \sim 1.4 r_{200}$ in a dark halo 
and is called a "splashback radius" \citep{Diemer2014,More2015,Deason2020}.
In particular, \citet{Diemer2014} and \citet{More2015} revealed, using cosmological N-body simulations,
that the splashhback radius falls in the range of
(0.8-1.0) $r_{\rm 200}$ for rapidly accreting halos and is $\sim 1.5 r_{\rm 200}$ for slowly accreting halos.
For Milky Way-mass galaxies, a sharp decrease in the density slope of the stellar halo, which called a "caustic radius", 
also appears in (0.3 - 0.8) $r_{\rm 200}$, which is associated with the second caustic radius of a dark halo
as stated in \citet{Deason2020}.

Therefore, to get insights into these characteristic radii from the current data, we attempt to fit the radial distribution
of our BHB sample to the mixture of the Einasto profile \citep{Einasto1965} for the inner part of the halo and the power-law profile for it outermost part
as defined in \citet{Diemer2014}:
\begin{eqnarray}
& &\rho(r_q) = f_{\rm trans}\rho_{\rm einasto}+\rho_{\rm outer} \\
& &\rho_{\rm einasto} = \rho_{\rm 0} \exp \Bigl( -\frac{2}{n}\Bigl[\Bigl( {\frac{r_q}{r_{\rm s}}} \Bigr)^n-1 \Bigr] \Bigr) \\
& &f_{\rm trans} = \Bigl[1+ \Bigl(\frac{r_q}{r_{\rm t}} \Bigr) ^\beta \Bigr) \Bigr]^{-\gamma/\beta} \\
& &\rho_{\rm outer} = \rho_{\rm m} \Big[ b_{\rm e} \Bigl(\frac{r_q}{5r_{\rm 200}} \Bigr)^{-S_{\rm e}}+1 \Bigr]\\
& &r_q^2 = x^2+y^2+z^2q^{-2} \ .
\label{eq:NFWEinasto}
\end{eqnarray}
where the Einasto profile, $\rho_{\rm einasto}$, is characterized by three parameters $(n,r_{\rm s},\rho_0)$, and the transition term,
$f_{\rm trans}$, provides the steepening of the profile at around a truncation radius, $r_{\rm t}$, where $\beta$ and $\gamma$
give the steepness of the profile at $r \sim r_{200}$. The outermost profile, $\rho_{\rm outer}$, is given by a power-law with an index $S_{\rm e}$
plus the mean density of the universe, $\rho_m$.

\begin{figure*}[t!]
\begin{center}
\includegraphics[width=80mm]{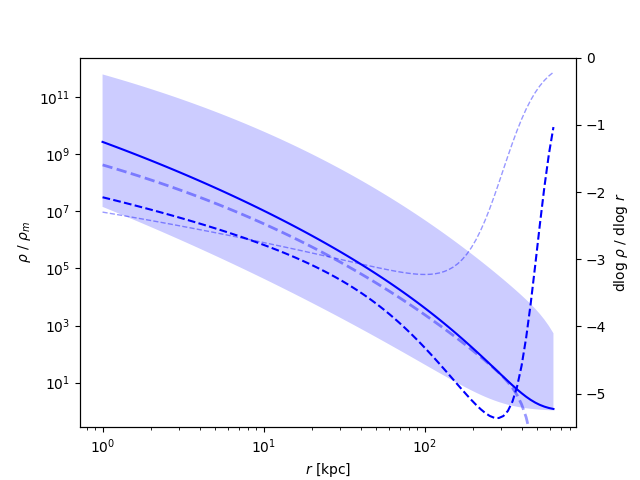}
\includegraphics[width=80mm]{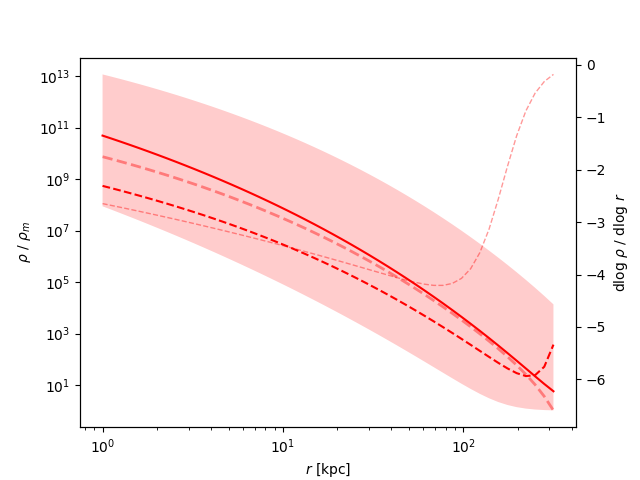}
\end{center}
\hspace*{10mm}
\caption{
The left (right) panel show the best fit results for BHBs (BSs).
The blue (red) solid line is the density distribution, and the light blue (light red) shadow shows the error range.
The blue (red) dashed blue line indicates the slope of the density distribution. The thick light blue (light red) dashed line shows the slope at the upper limit of the error, and the thin light blue (light red) dashed line shows the slope at the lower limit of the error.
}
\label{fig: einasto_profile}
\end{figure*}

The best-fit values for the parameters in this function form are summarized in Table \ref{tab:best_fit_einasto}
and the best-fit density profiles for BHBs ad BSs are shown in Figure \ref{fig: einasto_profile}.
For BHBs, a rapid increase in the slope between $r = 400$ kpc and 500 kpc is 
likely to be observed, but no clear results for a splashback/caustic radius are obtained from the current sample.
On the other hand, a sharp increase in the slope is observed between $r = 200$ kpc and 300 kpc for BSs.
This result is roughly consistent with the finding that the edge of the Milky Way is located at $r = 292$ kpc
obtained from the motion of dwarf galaxies in \citet{Deason2020}.
However, as is clear from Figure \ref{fig: einasto_profile}, the current result suffers from a rather large error,
so the derivation of a splashback/caustic radius is yet inconclusive in our work.

\begin{table*}
\tbl{Best fit parameters for Einasto model}{%
\begin{tabular}{l|l|l|l|ccc}
\hline
 BHB                    & BS              &
          $f_{\rm BHB}$ & $f_{\rm WD}$ & $f_{\rm QSO}$       \\
\hline\hline
$\log (\rho_{0}/\rho_{\rm m}) =9.7^{+1.2}_{-1.0}$, 
        & $\log (\rho_{0}/\rho_{\rm m}) =11.5^{+1.1}_{-1.2}$,  
        & $-0.210^{+0.032}_{-0.052}$ & $0.882^{+0.008}_{-0.007}$ & $0.279^{+0.007}_{-0.007}$   \\
        $\log (n)=-0.92^{+0.19}_{-0.25}$, 
        & $\log (n)=-0.81^{+0.16}_{-0.13}$,  & & & \\
        $\log(r_{\rm s})=-0.13^{+0.65}_{-0.58}$, $\log (b_{\rm e}) =9.9^{+0.9}_{-1.3}$,
        & $\log(r_{\rm s})=-0.35^{+0.60}_{-0.55}$, $\log (b_{\rm e}) =10.0^{+1.1}_{-1.0}$, & & & \\
        $S_{\rm e}=9.1^{+1.1}_{-1.0}$, $\log(r_{\rm t})=2.3^{+0.5}_{-0.4}$,
        & $S_{\rm e}=10.2^{+0.8}_{-0.7}$, $\log(r_{\rm t})=2.3^{+0.5}_{-0.5}$, & & & \\
        $\log(\beta)=0.14^{+0.79}_{-0.51}$, $\log(\gamma)=0.35^{+0.53}_{-0.68}$,
        & $\log(\beta)=-0.17^{+0.81}_{0.53}$, $\log(\gamma)=0.22^{+0.47}_{-0.64}$, & & & \\
        $\log(q)=0.12^{+0.13}_{-0.12}$  & $\log(q)=0.14^{+0.05}_{-0.06}$    & & &\\
\hline
\end{tabular} }
\label{tab:best_fit_einasto}
\end{table*}

\subsection{The effects of the photometric errors on the results}

\begin{figure*}[t!]
\begin{center}
\includegraphics[width=170mm]{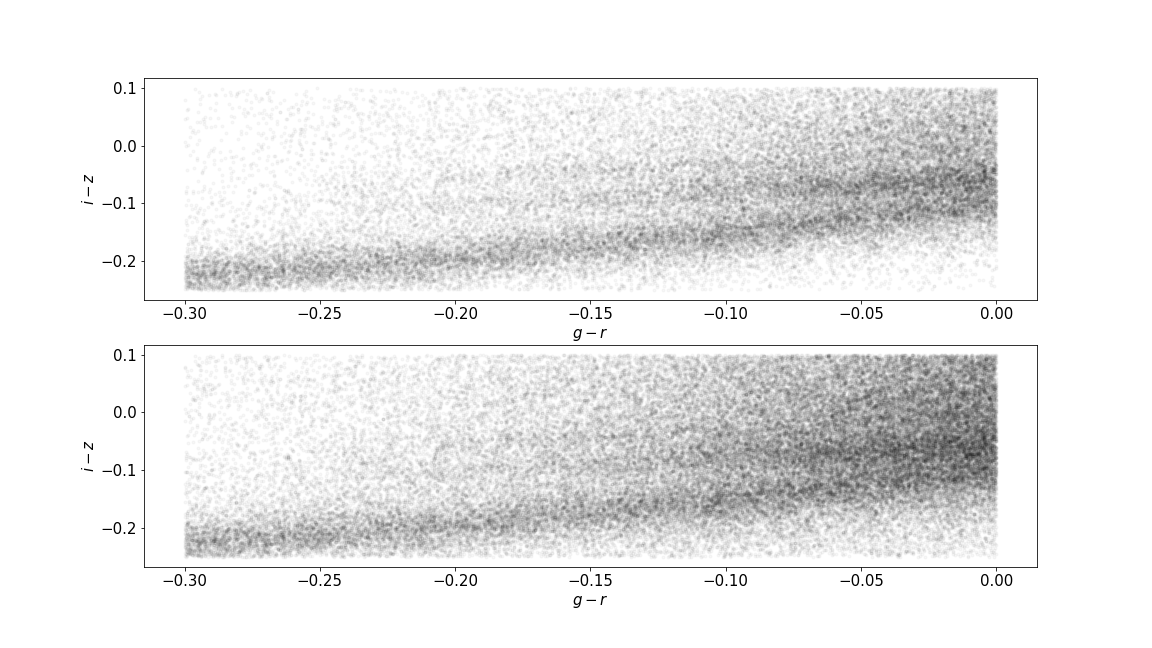}
\end{center}
\hspace*{10mm}
\caption{
The $g-r$ vs. $i-z$ diagram for S21A data by excluding the stars with g-band photometric errors exceeding $0.05$ mag (upper panel) and for all the stars (lower panel).
}
\label{fig: colormap}
\end{figure*}

First, to investigate how the photometric errors affect the color distribution
shown in Figure \ref{fig: color-color}, 
we compare the $g-r$ vs. $i-z$ distribution between the sample by excluding the data with errors larger than 0.05~mag
and that without this restriction in the analysis, as presented in Figure \ref{fig: colormap}.
It is clear that the color distribution of BHBs and BS remains the same even considering the photometric errors.

\begin{figure*}[t!]
\begin{center}
\includegraphics[width=170mm]{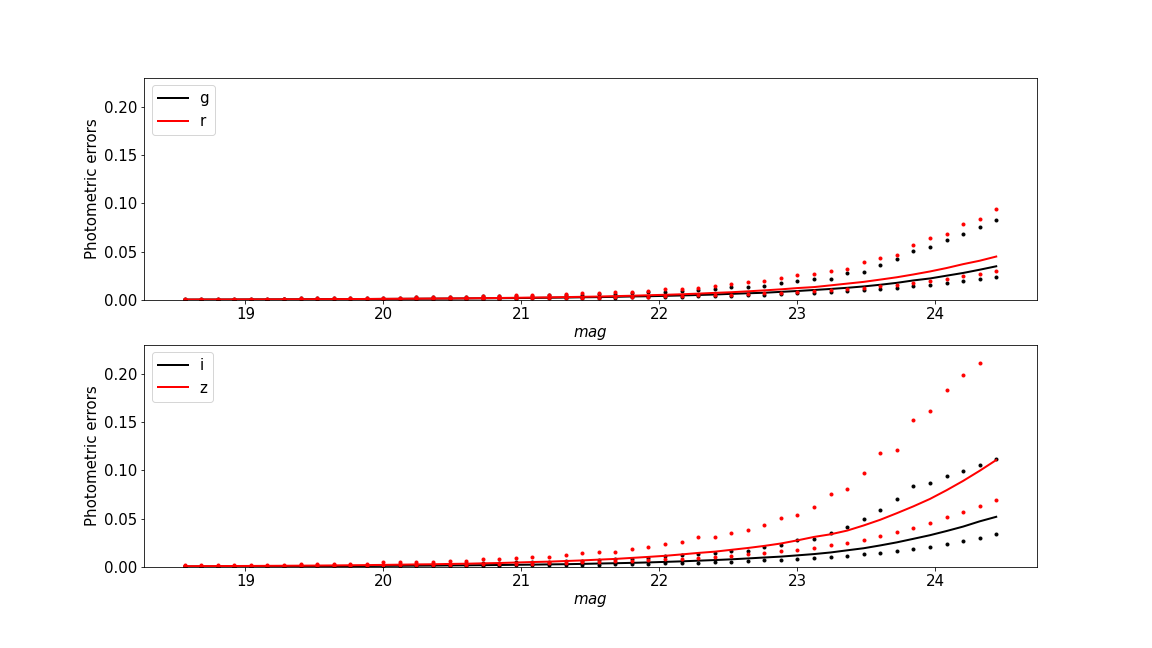}
\end{center}
\hspace*{10mm}
\caption{

Photometric errors in the current data.
The black and red lines in the upper panel are for the g-band and
r-band, respectively, and the black and red lines in the lower panel
are for the i-band and z-band, respectively.
The dotted lines correspond to the $5\%$ and $95\%$ percentile data for each band
and the median is shown as a solid line.
    
}
\label{fig: photometricerr}
\end{figure*}

Next, we perform an additional analysis using only the samples with magnitudes of $g < 23$
(i.e., within $r=300$ kpc) with the photometric errors smaller than 0.05~mag.
The photometric errors for each band are shown in Figure \ref{fig: photometricerr}.
As presented in Table \ref{tab:best_fit_under23}, we have obtained the similar results to those for the samples
without this restriction shown in Section \ref{sec:Best fit}, thereby suggesting that
the impact of the photometric errors on the results is limited. It is worth noting that
a discontinuous change of $\alpha$ at a break radius $r_{\rm b}$ is more pronounced for the samples
with low photometric errors than the samples without this restriction, although the obtained $r_{\rm b}$ is different.
The BIC value for ABPL is the largest, indicating that this is the optimal model.
This result is understood from the fact that excluding the data with large photometric errors leads to the lack of
the sample stars beyond $r=300$ kpc, that mitigates the effect of the tendency for $\alpha$ to become smaller again at
$r>300$ kpc. As mentioned above, the samples at such radii contain large photometric errors, suggesting that it is
necessary to compare these results with those available from future deeper surveys, as discussed in Section \ref{sec:MW edge}.



\begin{table*}
\tbl{Best fit parameters for sample $g$ under $23$ mag}{%
\begin{tabular}{l|l|l|cccc}
\hline
Model & BHB & BS & $f_{\rm BHB}$ & $f_{\rm WD}$ & $f_{\rm QSO}$ & $\Delta$BIC \\
\hline\hline
ASPL & $\alpha=4.51^{+0.26}_{-0.25}$, $q=1.51^{+0.04}_{-0.03}$ & $\alpha=4.63^{+0.10}_{-0.11}$, $q=1.40^{+0.02}_{-0.02}$ & $0.20^{+0.02}_{-0.02}$ & $0.86^{+0.01}_{-0.01}$ & $0.29^{+0.01}_{-0.01}$ & 76 \\
\hline
ABPL & $\alpha_{\rm in}=4.12^{+0.38}_{-0.37}$, $\alpha_{\rm out}=16.12^{+2.64}_{-4.75}$ & $\alpha_{\rm in}=4.35^{+0.13}_{-0.13}$, $\alpha_{\rm out}=16.11^{+2.51}_{-3.19}$ & $0.19^{+0.02}_{-0.02}$ & $0.86^{+0.01}_{-0.01}$ & $0.29^{+0.01}_{-0.01}$ & 0 \\
& $r_{\rm b}/{\rm kpc}=118^{+32}_{-18}$, $q=1.75^{+0.50}_{-0.32}$ & $r_{\rm b}/{\rm kpc}=55^{+5}_{-4}$, $q=1.29^{+0.11}_{-0.09}$ & & & & \\
\hline
ASPL & $\alpha=4.24^{+0.40}_{-0.28}$, $r_0/{\rm kpc}=177^{+52}_{-62}$ & $\alpha=4.28^{+0.32}_{-0.24}$, $r_0/{\rm kpc}=67^{+34}_{-20}$ & $0.20^{+0.03}_{-0.02}$ & $0.86^{+0.01}_{-0.01}$ &  $0.29^{+0.01}_{-0.01}$& 65 \\
with $q(r)$ & $q_{0}=1.94^{+0.42}_{-0.38}$, $q_{\infty}=0.52^{+0.62}_{-0.36}$ & $q_{0}=1.58^{+0.30}_{-0.21}$, $q_{\infty}=0.50^{+0.94}_{-0.37}$ & & & & \\
\hline
\end{tabular} }
\label{tab:best_fit_under23}
\end{table*}
\section{Conclusions}

We have selected and analyzed candidate BHBs from the HSC-SSP Wide layer data obtained until 2021 January (S21A), which covers $\sim 1,100$~deg$^2$ area, 
based on an extensive Bayesian method to minimize the effects of non-BHB contamination as much as possible.
In this selection method, the $z$-band brightness of a selected stellar target is useful as a probe of a surface 
gravity of a BHB star against other A-type stars. 

Applying our selection method to the sample with $18.5 < g < 24.5$, which, for candidate BHBs, corresponds 
to the Galactocentric radii at $r = 36 \sim 575$~kpc, 
we have obtained the density slopes of BHBs for a single power-law model as $\alpha = 
4.11^{+0.18}_{-0.18}$ and for a broken power-law model as $\alpha_{\rm 
in}=3.90^{+0.24}_{-0.30}$ and $\alpha_{\rm out}=9.1^{+6.8}_{-3.6}$ divided at a 
radius of $r_{\rm b}=184^{+118}_{-66}$~kpc. The difference in statistical significance between these power-law models
appears small. For thees models allowing a non-spherical halo shape, 
an axial ratio of $q=1.56^{+0.34}_{-0.23}$ or $q = 1.35^{+0.11}_{-0.19}$ corresponding to a prolate shape
is most likely. It is also suggested from the spatial distribution of th currently selected BHBs that
the MW stellar halo may have a trough at around $r \sim 200$~kpc,
although this needs to be assessed using the further survey data.

The density slope obtained in this work is basically in agreement with that from the 
CFIS survey for BHBs \citep{Thomas2018}. However, it is systematically 
shallower than the slope derived from RRLs at $r$ below $\sim 100$~kpc 
\citep{Cohen2017, Hernitschek2018}. This may be simply due to the different radial 
range of each sample, $r < 100$~kpc for RRLs and $50 < r < 360$~kpc for BHBs, or 
RRLs may have an intrinsically more centrally concentrated distribution than BHBs.
However, distant data contains at least some errors, 
and in order to understand the global structure of the MW's stellar halo, 
a further analysis using wider and deeper data is necessary such as with the LSST survey at the Vera Rubin Observatory
and the ongoing Ultraviolet Near-Infrared Northern Sky Survey (UNIONS) that covers the footprint of the Euclid Survey \citep{Euclid2022}\footnote{UNIONS combines multi-band photometric images from different telescopes: $u-$ and $r-$band images from CFHT called CFIS, $i-$ and $z-$band from Pan-STARRS, $z-$ and $g-$band from Subaru HSC called WISHES and WHIGS, respectively (WISHES: Wide Imaging with Subaru HSC of the Euclid Sky, WHIGS: Waterloo Hawai'i IFA Survey).}.
Also, to interpret such observational results in the form of the past 
merging history, more extensive numerical simulations for the formation of stellar halos 
will be important, where not only accretion/merging of satellites from outside but also 
the {\it in situ} formation of halo stars are properly taken into account.

\begin{ack}
This work is supported in part by JSPS Grant-in-Aid for Scientific 
Research (B) (No. 25287062) and MEXT Grant-in-Aid for Scientific Research
(No. 18H05437, 21H05448, and 24K00669 for MC,
No. JP20H01895, JP21K13909, and JP23H04009 for KH).
TM is supported by a Gliese Fellowship at the Zentrum f\"{u}r Astronomie, University of Heidelberg, Germany.

The Hyper Suprime-Cam (HSC) collaboration includes the astronomical
communities of Japan and Taiwan, and Princeton University.  The HSC
instrumentation and software were developed by the National
Astronomical Observatory of Japan (NAOJ), the Kavli Institute for the
Physics and Mathematics of the Universe (Kavli IPMU), the University
of Tokyo, the High Energy Accelerator Research Organization (KEK), the
Academia Sinica Institute for Astronomy and Astrophysics in Taiwan
(ASIAA), and Princeton University.  Funding was contributed by the FIRST 
program from Japanese Cabinet Office, the Ministry of Education, Culture, 
Sports, Science and Technology (MEXT), the Japan Society for the 
Promotion of Science (JSPS),  Japan Science and Technology Agency 
(JST),  the Toray Science  Foundation, NAOJ, Kavli IPMU, KEK, ASIAA,  
and Princeton University.
This paper makes use of software developed for the Large Synoptic Survey Telescope. We thank the
LSST Project for making their code freely available. The Pan-STARRS1 (PS1) Surveys have been made
possible through contributions of the Institute for Astronomy, the University of Hawaii,
the Pan-STARRS Project Office, the Max-Planck Society and its participating institutes, the Max Planck Institute for
Astronomy and the Max Planck Institute for Extraterrestrial Physics, The Johns Hopkins University,
Durham University, the University of Edinburgh, Queen's University Belfast, the Harvard-Smithsonian
Center for Astrophysics, the Las Cumbres Observatory Global Telescope Network Incorporated, the
National Central University of Taiwan, the Space Telescope Science Institute, the National Aeronautics
and Space Administration under Grant No. NNX08AR22G issued through the Planetary Science Division
of the NASA Science Mission Directorate, the National Science Foundation under Grant
No.AST-1238877, the University of Maryland, and Eotvos Lorand University (ELTE).
\end{ack}


\end{document}